\documentclass[aps,prb,showpacs,twocolumn,superscriptaddress]{revtex4-1}
\usepackage{graphicx}
\usepackage{braket}
\usepackage{amsmath}
\usepackage{amssymb}
\usepackage{color}
\usepackage{soul}

\usepackage{notes2bib}

\begin{document}

%\preprint{}

\title{Localization landscape theory of disorder in semiconductors III: Application to carrier transport and recombination in light emitting diodes}


\author{Chi-Kang Li}
\affiliation{Graduate Institute of Photonics and Optoelectronics and Department of Electrical Engineering, National Taiwan University, Taipei 10617, Taiwan}

\author{Marco Piccardo}
%\email[]{marco.piccardo@polytechnique.edu}
\affiliation{Laboratoire de Physique de la Mati\`ere Condens\'ee, Ecole polytechnique, CNRS, Universit\'e Paris Saclay, 91128 Palaiseau Cedex, France}

\author{Li-Shuo Lu}
\affiliation{Graduate Institute of Photonics and Optoelectronics and Department of Electrical Engineering, National Taiwan University, Taipei 10617, Taiwan}

\author{Svitlana Mayboroda}
%\email[]{svitlana@math.umn.edu}
\affiliation{School of Mathematics, University of Minnesota, Minneapolis, Minnesota 55455, USA}

\author{Lucio Martinelli}
%\email[]{lucio.martinelli@polytechnique.edu}
\affiliation{Laboratoire de Physique de la Mati\`ere Condens\'ee, Ecole polytechnique, CNRS, Universit\'e Paris Saclay, 91128 Palaiseau Cedex, France}

\author{Jacques Peretti}
%\email[]{jacques.peretti@polytechnique.edu}
\affiliation{Laboratoire de Physique de la Mati\`ere Condens\'ee, Ecole polytechnique, CNRS, Universit\'e Paris Saclay, 91128 Palaiseau Cedex, France}

\author{James S. Speck}
\affiliation{Materials Department, University of California, Santa Barbara, California 93106, USA}

\author{Claude Weisbuch}
%\email[]{claude.weisbuch@polytechnique.edu}
\affiliation{Laboratoire de Physique de la Mati\`ere Condens\'ee, Ecole polytechnique, CNRS, Universit\'e Paris Saclay, 91128 Palaiseau Cedex, France}
\affiliation{Materials Department, University of California, Santa Barbara, California 93106, USA}

\author{Marcel Filoche}
%\email[]{marcel.filoche@polytechnique.edu}
%\thanks{}
\affiliation{Laboratoire de Physique de la Mati\`ere Condens\'ee, Ecole polytechnique, CNRS, Universit\'e Paris Saclay, 91128 Palaiseau Cedex, France}

\author{Yuh-Renn Wu}
\email[]{yrwu@ntu.edu.tw}
\affiliation{Graduate Institute of Photonics and Optoelectronics and Department of Electrical Engineering, National Taiwan University, Taipei 10617, Taiwan}

\date{\today}

\begin{abstract}
This paper introduces a novel method to account for quantum disorder effects into the classical drift-diffusion model of semiconductor transport through the localization landscape theory. Quantum confinement and quantum tunneling in the disordered system change dramatically the energy barriers acting on the perpendicular transport of heterostructures. In addition they lead to percolative transport through paths of minimal energy in the 2D landscape of disordered energies of multiple 2D quantum wells. This model solves the carrier dynamics with quantum effects self-consistently and provides a computationally much faster solver when compared with the Schr\"odinger equation resolution. The theory also provides a good approximation to the density of states for the disordered system over the full range of energies required to account for transport at room-temperature. The current-voltage characteristics modeled by 3-D simulation of a full nitride-based light-emitting diode (LED) structure with compositional material fluctuations closely match the experimental behavior of high quality blue LEDs. The model allows also a fine analysis of the quantum effects involved in carrier transport through such complex heterostructures. Finally, details of carrier population and recombination in the different quantum wells are given.
\end{abstract}

\pacs{71.23.An, % Electronic structure of disordered solids: Theories and models; localized states
72.15.Rn, % Electronic conduction in metals and alloys: Localization effects (Anderson or weak localization),
03.65.Ge % Quantum Mechanics: Solutions of wave equations: bound states
%73.22.Dj % Electronic structure of nanoscale materials and related systems: Single particle states
}
%\keywords{}

\maketitle

\section{Introduction}

In a previous paper (referred to here as LL1, Ref.~\citenum{Filoche2017}) we have introduced the new localization landscape (LL) theory to describe energy levels, localized states and density of states in disordered materials, with applications to the nitride materials case. A second paper~ (refered to here as LL2, Ref.~\citenum{Piccardo2017}) showed the capability of the theory to compute materials system for time independent, frozen states entering optical absorption below the bandgap in alloy quantum wells (QWs). This phenomenon called the Urbach tail was shown to be well described by the density of states and the wavefunction overlap given by the landscape theory. The present paper will address the physical effects of disorder on carrier transport and recombination in a much more complex situation, that of the full multilayer heterostructures which make the light emitting diodes (LEDs), devices of utmost importance for energy savings and so far very poorly modeled due to the lack of a proper description of the effects of disorder.

Applying the LL theory to nitride-based alloys aims at solving long standing issues in these important materials. In recent years, nitride-based materials indeed play an increasing role in semiconductor markets including high power devices and light emitters due to the large bandgap range and full visible spectrum achieved by nitride ternary alloys.\cite{Weisbuch2015, Nakamura1991, Pengelly2012} However, a complete fundamental analysis of the intrinsic material properties is still missing to reconcile experiments and theoretical calculations. For instance, the explanations for the droop effect in nitride-based LEDs include electron overflow,\cite{Kim2007, Ozgur2011} Auger recombination,\cite{Iveland2013, Shen2007, Binder2013a, David2010b} poor hole injection,\cite{Ni2008} and carrier localization-delocalization.\cite{Mukai1999, WatsonParris2011} While various direct measurements favor the Auger recombination, however requiring a large Auger coefficient when considering direct Auger recombination processes,\cite{Kioupakis2015} a number of other indirect experiments analyzed with simple or ad-hoc models form the basis for claiming the other mechanisms. All these above mechanisms are probably influenced by material compositional disorder effects which are neglected in device simulations or are only included at an elementary level. The disorder is best described from experiments including atom probe tomography (APT) with indium atoms distributed randomly in InGaN/GaN quantum wells (QW).\cite{WatsonParris2011, Bennett2011, Shivaraman2013, Mazumder2013, Riley2014} Such compositional disorder due to local indium fluctuations should have an important role in determining electrical and optical properties in nitride-based LEDs or laser diodes. Recent atomistic calculations show that the localization effect induced by indium fluctuations will cause strong inhomogeneous broadening of the lowest transition energies.\cite{Schulz2015} Radiative recombination coefficients were also found to decrease with increasing indium concentration due to increased carrier localization in the random alloy fluctuations.\cite{AufderMaur2016} However, atomistic simulations are impractical for calculating a full LED structure with contacts, multiple quantum wells (MQWs), electron blocking layer (EBL), and indium fluctuations. They do not really allow computing basic device characteristics, such as current-voltage curves and electro-luminescence spectra, where the determination of many quantum levels and the description of transport among such states is required. 

In fully ordered materials, e.g., pure compounds, because of the small computation time and of the mature development of the technique, the classical drift-diffusion (DD) equations coupled with the Poisson equation are widely used to describe transport and optical properties although they treat carriers as semi-particles with a renormalized effective mass. Many results of quantum theory are however implicitly introduced as energy levels, density of states, quantum Fermi-Dirac statistics, and transport parameters, such as carrier mobilities and diffusion coefficients. In the case of LED simulations, radiative and non-radiative recombination mechanisms are described by a Shockley-Read-Hall coefficient~$A$, a radiative coefficient~$B$, and a non-radiative Auger recombination coefficient~$C$. However, when comparing computational results using typical material parameters for ideal quantum wells or barriers without compositional disorder to experiments, the classical transport model leads to turn-on voltages either much too large in LEDs~\cite{Yang2014, Karpov2011, Chen2009} or too small in electron barriers.\cite{Browne2015} It also does not satisfactorily model the droop of the internal quantum efficiency (IQE).\cite{Piprek2010} In general, researchers then use reduced piezo-electric polarization charge and larger Auger coefficient compared with experimental measurements to fit the experimental quantum efficiency~\cite{Piprek2010} and the turn-on voltage.\cite{Yang2014, Wu2012}

Quantum models such as the non-equilibrium Green’s function formalism (NEGF) have been proposed to model quantum transport properties~
.\cite{Shedbalkar2016, Datta1995} They provide a good description of quantum effects such as tunneling which are absent from DD models. However, to describe indium fluctuations~\cite{Wu2015} multi-dimensional models are needed and the large burden of computation time ($>$ thousands of CPU hours) makes the NEGF approach impractical. 

We previously incorporated indium fluctuations into the classical Poisson and DD model, by taking into account the fluctuating conduction and valence band potentials. We found that inclusion of these random potentials led to the enhancement of Auger recombination due to the higher local carrier density. In addition, the many percolation paths through local minima of potential energy for carrier diffusion did reduce the LED turn-on voltage.\cite{Yang2014, Wu2012} It was also computed that carrier localization induced by indium fluctuations has a strong influence on the broadening of the light emission spectrum.\cite{Yang2014} However, we still did not take into account the wave nature of electrons leading to localization and delocalization in a disordered potential. 

In this paper, we will implement the LL~theory~\cite{Filoche2012, Arnold2016} into semiclassical Poisson and drift-diffusion equations. We model carrier dynamics including transport and recombination by using known parameters (mobilities, $A$, $B$, and $C$ coefficients) from experiments. The quantum effects affecting in-plane and perpendicular transport which arise from indium fluctuations are taken into account by effective electron and hole energies (the landscape energies), electron-hole overlap, disordered densities of states. Thanks to the efficiency of landscape computations, we carry out all calculations self-consistently for the disordered LED system.

\section{Simulation Methods}

In this section, we describe the simulation framework, including how to apply the LL~theory into Poisson and DD equations. In the standard classical picture, Poisson and drift-diffusion equations are solved self-consistently to obtain the conduction and valence band edges which are the potential energies $E_{c,v}$ for electrons and holes. The set of equations is
\begin{align}\label{eq:Poisson-DD}
\begin{cases}
\nabla \cdot \left(\varepsilon \nabla \varphi \right) = e \left(n - p + N_A^- - N_D^+ \pm \rho_{pol} \right)\\
J_n = n\mu_n \nabla E_{Fn}\\
J_p = n\mu_p \nabla E_{Fp}\\
\nabla \cdot J_{n,p} = \pm e \left( A_0 + B_0 np + C_0 \left(n^2p + np^2\right)\right)\\
A_0 = \frac{np-n_i^2}{\tau_n \left(p + n_i e^\frac{\left(E_i-E_t\right)}{k_BT}\right) + \tau_p \left(n + n_i e^\frac{\left(E_t-E_i\right)}{k_BT}\right)}\\
n = \displaystyle \int_{E_c}^{+\infty} {\rm DOS}_{n,bulk}(E) \cdot f_n(E)~dE\\
p = \displaystyle \int_{-\infty}^{E_v} {\rm DOS}_{p,bulk}(E) \cdot f_p(E)~dE\\
\end{cases}
\end{align}
where $\varphi$ is the electrostatic potential, $E_{Fn}$ and $E_{Fp}$ are the  quasi Fermi levels of electrons and holes, DOS$_{n,p,bulk}(E)=\sqrt{|E-E_{c,v}|} \cdot \sqrt{2} m^{*3/2}/\left(\pi^2 \hbar^3\right)$ is the bulk density of states (DOS), $m^*$ is the effective mass of the electron or hole, and $f_n$ and $f_p$ are the Fermi distribution functions for electrons and holes, respectively. Note that if all the layers of the simulated semiconductor structure are considered as homogeneous, then the local free electron and hole carrier densities, $n$ and $p$, will only depend on the growth direction $z$. $E_c$ and $E_v$ are the local conduction and valence band potential energies, respectively. $N_A^-$ and $N_D^+$ are the activated doping densities of acceptors and donors, respectively. $\rho_{pol}$ is the density of polarization charges, which can be computed by taking the divergence of the total polarization in the space $(\nabla \cdot \bf{P}^{\rm total})$ including spontaneous and piezoelectric polarization fields. $J_n$ and $J_p$ are the electron and hole current densities, respectively. This paper uses the Shockley-Read-Hall (SRH) model to account for the defect-related non-radiative (NR) recombination through a rate $A_0$ [Eq.~(\ref{eq:Poisson-DD})], where $\tau_n$ and $\tau_p$ are the NR carrier lifetimes dependent on the growth condition, which are taken in this paper as $10^{−7}$~s (300~K), a value typically associated with low-defect density nitride LEDs.\cite{DeSanti2016, Zhang2009} $k_B$~is the Boltzmann constant and $T$ is taken as room temperature. $E_t$ is the trapping energy level assumed to be located at the mid-gap, and $E_i$ and $n_i$ are the intrinsic energy level and intrinsic carrier density, respectively. $B_0$~($3.0 \times 10^{−11}$~cm$^3$s$^{-1}$) is the intrinsic radiative recombination coefficient. Our $B_0 np$ rate represents the usual radiative recombination rate with $B$ including the effect of overlap of the wave functions across the active region.\cite{David2010a} In contrast, $B_0$ is a bulk coefficient in our model and the separations of the electron and hole distributions in the QW both along z due to the quantum-confined Stark effect (QCSE) and in-plane due to indium fluctuations are considered in the $np$ term. $C_0$ ($2.0 \times 10^{−31}$~cm$^6$s$^{-1}$, see Ref.~\citenum{Kioupakis2015}) is the Auger recombination coefficient, where the influence of the electron and hole overlap is included in the $n^2p$ and $np^2$ terms. $B_0$ and $C_0$ are considered temperature-independent in our simulations.

At this point we should emphasize that the choices of $B_0$ and $C_0$ are somewhat arbitrary if even not inconsistent: as the computation takes into account disorder induced localization effects and QW confinement Stark effects, we should take values for the $B_0$ and $C_0$ bulk parameters without disorder and electric field, but any experimental value will incorporate such effects of disorder and QW confinement. Only those values for bulk binary compounds such as GaN would be disorder effect free, but then the effect of the bandgap change with QW alloying might be significant. For $B_0$, Kioupakis et al.~\cite{Kioupakis2013} calculated that the change from GaN to In$_{0.25}$Ga$_{0.75}$N modifies $B_0$ by less than 10\% without taking disorder into account, with $B_0$=$6.4 \times 10^{-11}$~cm$^3$s$^{-1}$ for 25\% indium. On the experimental side, the bulk $B_0$ value ($7.0 \times 10^{−11}$~cm$^3$s$^{-1}$) extracted from InGaN/GaN double heterostructure (DH) LED experiments would give a value without QCSE, although with disorder, but it assumes that $n$ (carrier density) is constant across the structure.\cite{David2010c} However, in real DHs, the carrier density is not a constant as a large polarization electric field is present in the DH. A very high carrier density still exists locally to screen the polarization field in the DH which locally increases the radiative recombination rate compared to the space-averaged one. Therefore, the $B_0$ value obtained in a DH while assuming uniform carrier concentration is not correct and cannot be taken as the $B_0$ bulk input parameter without disorder and electric field as it is an overestimation. Thus, we rather take an experimental value of $3.0\times 10^{-11}$~cm$^3$s$^{-1}$ for $B_0$~\cite{David2010a} in QWs as a better approximation in this simulation, although a parametric evaluation should be done at a later point, both experimentally and theoretically.

The $C_0$ issue is even more complex: the Auger recombination term $Cn^3$ (more precisely $C_{eeh}n^2p + C_{ehh}np^2$, where $C_{eeh}$ and $C_{ehh}$ are the electron-electron-hole and electron-hole-hole Auger coefficients, respectively) incorporates disorder impact through two effects: increase in local carrier densities through carrier localization; increase in the Auger coefficient through wave function localization.\cite{Vaxenburg2013} The starting $C_0$ value for the computation should then be free of both effects, and cannot come from experiment. Theory~\cite{Kioupakis2015} gives a value of $\sim 10^{-33}$~cm$^6$s$^{-1}$ for the direct Auger process in InGaN and $\sim 10^{-31}$~cm$^6$s$^{-1}$ for the indirect phonon-assisted Auger process. By taking alloy effects through a supercell model, Kioupakis et al.~\cite{Kioupakis2015} calculate an alloy-assisted Auger coefficient of a few $10^{-31}$~cm$^6$s$^{-1}$, and an overall Auger coefficient of $\sim 10^{-30}$~cm$^6$s$^{-1}$. As we have not yet calculated the effect of alloy disorder on the Auger coefficient in the landscape model,\cite{Filoche2017, Lentali2017} we consider as a starting value for $C_0$ a low indium content experimental value from David and Grundmann~\cite{David2010a} of $\sim 2.0 \times 10^{−31}$~cm$^6$s$^{-1}$. Doing this, we might underestimate the effect of the alloying on the Auger coefficient, but we will still capture the major effect of local carrier concentration increase due to alloying, as well as the effects of the internal electric field.

If we want to solve the Schr\"odinger equation to take the disorder-induced quantum effects into account, we should use the disordered potential energy in a Schr\"odinger solver to calculate the wave functions and eigen-energies. Then the carrier density distribution can be obtained by the wave function distribution and relative eigen-energy levels. When the carrier density is obtained, this should be plugged into the Poisson-DD solver and the corresponding equations should be solved iteratively until convergence. In addition, a self-consistent 3D Poisson-Schr\"odinger solver is highly time consuming, as will be discussed later in Appendix.

As a result, we rather apply the theoretical landscape model proposed by Filoche et al.~\cite{Filoche2012, Arnold2016, Filoche2017} to obtain the equivalent semiclassical confining potential seen by the carriers, as described in LL1 (Ref.~\citenum{Filoche2017}). According to this theory, an effective potential can be found which captures the complex interference pattern created by the carrier wave functions in the original disordered potential and transforms it into a semiclassical confining potential which localizes the carriers in different regions. Additionally, the long-range exponential decay, characteristic of Anderson localization,\cite{Anderson1958} is explained as the consequence of multiple tunneling in the dense network of barriers created by this effective potential.\cite{Arnold2016} Therefore, both quantum localization/confinement and tunneling effects are described in the LL~theory.

\begin{figure}
\includegraphics[width=0.45\textwidth]{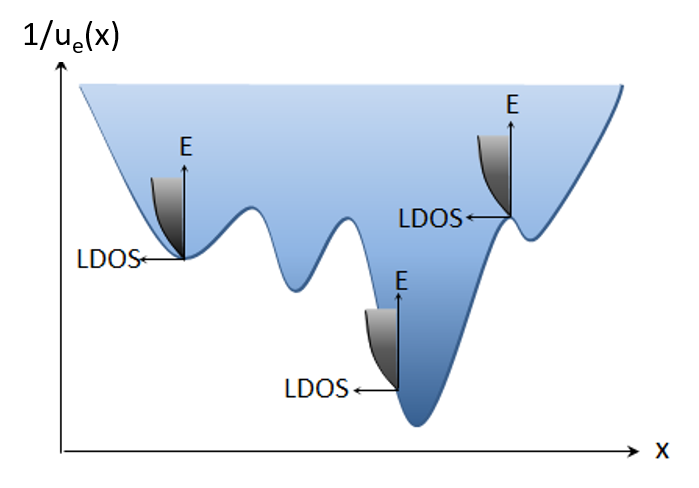}
\caption{Schematic of the local density of states (LDOS) arising from the landscape potential 1/ue for electrons. For simplicity the effective potential is shown in 1D.}
\label{fig:LDOS}
\end{figure}

The Hamiltonian entering the Schr\"odinger equations for electrons and holes reads
\begin{equation}
\widehat{H} = -\frac{\hbar^2}{2m^*_{e,h}} \Delta + E_{c,v}
\end{equation}
The landscapes $u_{e,h}(\vec{r})$ for electrons and holes are defined as the solutions of:
\begin{equation}\label{eq:landscape}
\widehat{H}~u_{e,h}\left(\vec{r}\right) = 1~,
\end{equation}
and $1/u_{e,h}$ are the effective potentials incorporating the localization properties of the solutions of the Schr\"odinger equation.\cite{Filoche2017, Arnold2016} The boundary conditions for Eq.~(\ref{eq:landscape}) can be either Dirichlet, Neumann or periodic. After obtaining $1/u_e$ and $1/u_h$ for electrons and holes, respectively, we use these as the input potential energies for the Poisson and DD equations to replace the original terms $E_c$ and $E_v$. In the carrier density calculation, $1/u_e$ and $1/u_h$ determine locally the bottom energy for the local DOS (LDOS) of the disordered system~(Fig.~\ref{fig:LDOS}):
\begin{align}\label{eq:carrier_densities}
\begin{aligned}
n = \displaystyle \int_{1/u_e}^{+\infty} {\rm LDOS}_{3D}(E) \cdot \displaystyle \frac{1}{1 + \exp\left(\frac{E-E_{Fn}}{k_BT}\right)}~dE\\
p = \displaystyle \int_{-\infty}^{1/u_h} {\rm LDOS}_{3D}(E) \cdot \displaystyle \frac{1}{1 + \exp\left(\frac{E_{Fp}-E}{k_BT}\right)}~dE
\end{aligned}
\end{align}
where $E_{Fn}$ and $E_{Fp}$ are the quasi Fermi energies for electrons and holes, respectively.

The LDOS in landscape theory can be computed from Weyl’s law in 3D:\cite{Arnold2016, Filoche2017}
\begin{equation}\label{eq:LDOS_3D}
{\rm LDOS}_{3D}(E) = \frac{\sqrt{2}m^*_{e,h}}{\pi^2 \hbar^3} \sqrt{|E-1/u_{e,h}|}
\end{equation} 
The fact that the DOS based on the LL is simply obtained by replacing the original potentials $E_{c,v}$ with the effective potentials $1/u_{e,h}$ makes the LL~theory easily implementable into the classical Poisson and DD model. It should be emphasized at this point that Eq.~(\ref{eq:carrier_densities}) and (\ref{eq:LDOS_3D}) will give a good description of optical and transport properties, much better than through the use of a model with disorderless QWs, only because the LDOS spectrum is well described in the landscape theory as was shown in LL1 (Ref.~\citenum{Filoche2017}), section IVB.2.

\begin{figure}
\includegraphics[width=0.28\textwidth]{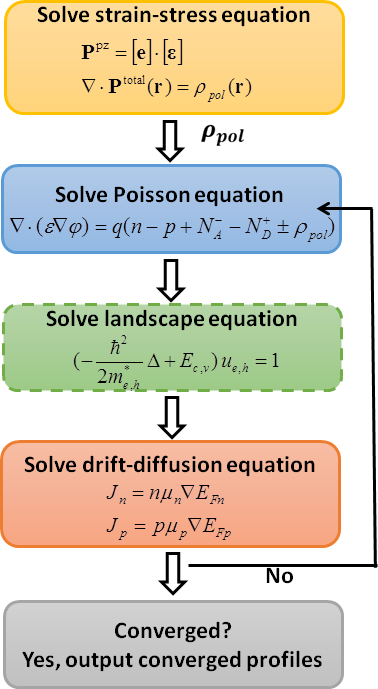}
\caption{Flowchart of Poisson and drift-diffusion equations by applying the LL~theory.}
\label{fig:flowchart}
\end{figure}

The schematic flowchart of the entire simulation process is shown in Fig.~\ref{fig:flowchart}. First, the spontaneous polarization charges and piezoelectric fields are computed. After the Poisson equation is solved to obtain $E_c$ and  $E_v$, the landscape equations for the conduction and valence band are solved giving the effective potentials $1/u_e$ and $1/u_h$ [Eq.~(\ref{eq:landscape})]. Then, the carrier densities of electrons and holes are calculated from Eq.~(\ref{eq:carrier_densities}) using $1/u_e$ and $1/u_h$, and fed back to the Poisson-DD equations to be solved in a self-consistent manner. When the potential energy difference between two consecutive iterations is smaller than $10^{-5}$~eV, we consider the simulation loop as having converged and the iterations stop.

\begin{figure}
\includegraphics[width=0.45\textwidth]{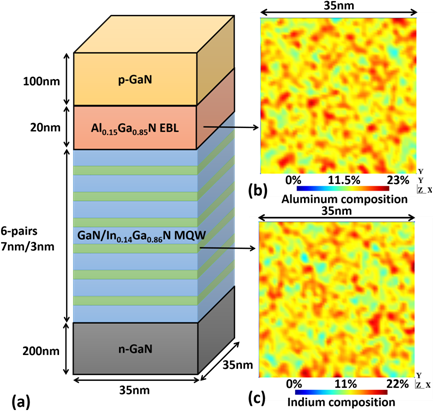}
\caption{(a)~Schematic full LED structure. (b)~In-plane aluminum distribution in the Al$_{0.15}$GaN$_{0.85}$N EBL layer. (c)~In-plane indium distribution in the third In$_{0.14}$GaN$_{0.86}$N QW layers. The Al and In distributions are generated by random numbers.}
\label{fig:schematic_LED}
\end{figure}

In the following sections, we model realistic GaN-based LED structures with indium fluctuations to study the impact of disorder as calculated through the LL~theory.

\section{The fluctuating potential in InGaN QWs and AlGaN EBL}

We adopt In$_{0.14}$Ga$_{0.86}$N and Al$_{0.15}$Ga$_{0.85}$N as the average alloy composition of the QWs and EBL, whose dimensions are illustrated in Fig.~\ref{fig:schematic_LED}(a). The indium and aluminum atoms are randomly distributed in the QWs and EBL and the composition maps shown in Fig.~\ref{fig:schematic_LED}(b) and \ref{fig:schematic_LED}(c) are obtained via the Gaussian averaging method detailed below.

Note that other alloy systems, for instance In$_x$Ga$_{1-x}$As, are well described by models that do not capture any effect of disorder-induced localization, such as the virtual crystal approximation (VCA) in which each potentially disordered site is substituted by an artificial atom interpolating between the properties of the actual components. In the VCA the maps of Fig.~\ref{fig:schematic_LED}(b) and \ref{fig:schematic_LED}(c) would be substituted by an homogeneous atom distribution with the mean composition of the alloy. However, nitride-based materials are characterized by composition fluctuations which induce large polarization-related local electric fields, large band offsets between GaN and InN, and heavy carrier effective masses, both larger than in the arsenide alloy system. Simulations based on the VCA in this context fail to provide a correct description of the local variations of the physical observables (density of states, carrier distribution, etc.).\cite{Nguyen2004}

Then one has to define the maximum length scale over which the rapidly oscillating distribution of atoms can be averaged to obtain a continuous fluctuating potential while preserving the effects of disorder on the electronic properties of the system. Such length scale can be predicted by the LL~theory, which is able to provide the effective potential fluctuations ``seen'' by the carriers. However, to avoid using a circular argument, let us estimate this length scale from an independent general theory of disorder: Baranovskii et al.~\cite{Baranovskii1978} showed that the spatial scale of the fluctuations affecting the transport of the electrons and holes is given by the de Broglie wavelength $\lambda = \hbar/\sqrt{m^* E_0}$, where $m^*$ is the carrier effective mass and $E_0$ is the energy scale of the band edge broadening due to disorder. In LL2 (Ref.~\citenum{Piccardo2017}) a value of $E_0\approx 50$~meV or $\approx30$~meV (fluctuations of $E_c$ and $E_v$ respectively) was calculated for In$_x$Ga$_{1-x}$N QW layers over a large range of indium concentrations ($x$=10\%-30\%). Taking these values of $E_0$ and the carrier effective masses of GaN (Table~\ref{tab:parameter-band}, note that $m^*_{\rm GaN} > m^*_{\rm InGaN}$) gives a lower bound on the spatial size of fluctuations in InGaN alloys of 2.8~nm for electrons and 1.1~nm for heavy holes, of the same order as the fluctuations of the effective confining potentials obtained from the LL~theory, as it will be shown in Fig.~\ref{fig:potential_maps}(b) and \ref{fig:potential_maps}(d). In the computational framework presented here the atomic disorder is smoothed by a Gaussian averaging over a length scale of $2\sigma \approx1.2$~nm, that is smaller than the spatial size of fluctuations seen by the carriers calculated from Ref.~\citenum{Baranovskii1978}. Therefore the averaged atomic distribution we use still incorporates disorder on the relevant scale for carrier transport. The details of the algorithm generating the electric potential map from the atomic distribution are described in the following.

\begin{figure}
\includegraphics[width=0.35\textwidth]{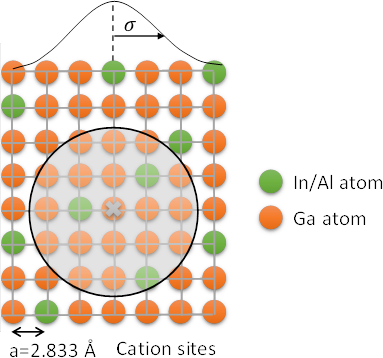}
\caption{The In, Al, and Ga atoms at each cation lattice site are assigned randomly by a random number generator. This possibility to obtain each atom is decided by the average alloy composition. The local composition at each atom site is determined by the Gaussian averaging method. If the mesh node does not coincide with the atom grid position, the linear interpolation of grid map will be used to determine the composition.}
\label{fig:sigma}
\end{figure}

At first, a cubic grid is constructed with a spacing corresponding to the average distance between cation atoms in GaN ($a = 2.833$~\AA). Then we randomly assign at each cation site either an indium (aluminum) or gallium atom for the InGaN (AlGaN) alloy, as shown in Fig.~\ref{fig:sigma}. For each atom site~$i$ the local averaged alloy composition $x(r_i)$ is determined from the Gaussian averaging method as
\begin{equation}
x\left(r_i\right) = \frac{\displaystyle \sum_j \mbox{atom}(j) \times e^{-\displaystyle\frac{\left(r_j-r_i\right)^2}{2\sigma^2}}}{\displaystyle \sum_j e^{-\displaystyle\frac{\left(r_j-r_i\right)^2}{2\sigma^2}}}\qquad,
\end{equation}
where the sum goes over all atom sites $j$ of the domain, $atom(j)$  is zero or unity as decided by the random generator, and $\sigma$ is the half width of the Gaussian broadening parameter.

In this paper and in LL2 (Ref.~\citenum{Piccardo2017}) we fixed $\sigma=2a$ ($\approx$ 0.6~nm), which gives an average alloy composition along the growth direction of an InGaN QW that matches APT data.\cite{Wu2012} In addition, we observe that when $\sigma=a$ is used, the indium composition map exhibits very large fluctuations, from 0\% to more than 60\%, that are strongly localized in space and behave like single-atom fluctuations, which are beyond the applicability of the effective-mass Schr\"odinger equation. Such choice of the Gaussian parameter impacts the results of the Poisson-drift-diffusion solver, which takes as an input the strongly fluctuating real potentials $E_c$ and $E_v$. However the calculated effective potentials used in the Poisson-drift-diffusion-landscape model are observed to be substantially unchanged for $\sigma$ values ranging from $a$ to $2a$, as the rapid fluctuations of the real potentials are smoothed by the LL~theory.

In our computations, we separate the process of atom grid generation and computation mesh construction to make the random alloy generator independent from the mesh elements. At each mesh node all the material parameters used in the simulation (e.g., bandgap, dielectric constant, and effective mass) are assigned according to the local alloy composition map $x(r_i)$. If the mesh node does not coincide with the atom grid position, the linear interpolation of grid map will be used to determine the composition. The III-nitrides material parameters we used are shown in Table~\ref{tab:parameter-band}. All parameters of InGaN and AlGaN alloys are obtained by an interpolation method as
\begin{equation}
\begin{aligned}
E_g^{\mbox{In}_x\mbox{Ga}_{1-x}\mbox{N}} = &~ (1-x)~E_g^{\mbox{GaN}} + x~E_g^{\mbox{InN}} \\& \quad - 1.4~x(1-x),\\
E_g^{\mbox{Al}_x\mbox{Ga}_{1-x}\mbox{N}} = &~ (1-x)~E_g^{\mbox{GaN}} + x~E_g^{\mbox{AlN}} \\& \quad - 0.8~x(1-x),\\
\varepsilon_r^{\mbox{In}_x\mbox{Ga}_{1-x}\mbox{N}} = &~(1-x)~\varepsilon_r^{\mbox{GaN}} + x~\varepsilon_r^{\mbox{InN}},\\
\varepsilon_r^{\mbox{Al}_x\mbox{Ga}_{1-x}\mbox{N}} = &~(1-x)~\varepsilon_r^{\mbox{GaN}} + x~\varepsilon_r^{\mbox{AlN}},\\
m^{*,\mbox{In}_x\mbox{Ga}_{1-x}\mbox{N}} = &~\left((1-x)/m^{*,\mbox{GaN}} + x/m^{*,\mbox{InN}}\right)^{-1},\\
m^{*,\mbox{Al}_x\mbox{Ga}_{1-x}\mbox{N}} = &~\left((1-x)/m^{*,\mbox{GaN}} + x/m^{*,\mbox{AlN}}\right)^{-1}
\end{aligned}
\end{equation}
The band offsets between GaN/InGaN and GaN/AlGaN conduction bands are assumed to be 63\% of the bandgap difference.

\begin{table}[tb!]
\begin{center}
\begin{tabular}{ccccccc}
\hline 
\hline
& $E_g$  & $\varepsilon_r$ & $m_{e}^{\parallel}$ & $m_{e}^{\perp}$ & $m_{hh}$ & $m_{lh}$ \\
~units~ & (eV) &  & ($m_0$) & ($m_0$) & ($m_0$) & ($m_0$)
\\ \hline
~GaN~  & 3.437 & 10.4  & 0.21 & 0.20 & 1.87 & 0.14 \\
~InN~  & 0.61  & 15.3  & 0.07 & 0.07 & 1.61 & 0.11 \\
~AlN~  & 6.0   & 10.31 & 0.32 & 0.30 & 2.68 & 0.26 \\
\hline
\multicolumn{4}{c}{Bandgap alloy} & \multicolumn{3}{c}{InGaN: 1.4} \\
\multicolumn{4}{c}{bowing parameter} & \multicolumn{3}{c}{AlGaN: 0.8} \\
\hline
\hline
\end{tabular}
\end{center}
\caption{Band structure parameters for wurtzite nitride alloys:\cite{Vurgaftman2001, Piprek2007} bandgap, relative permittivity, and effective masses.}
\label{tab:parameter-band}
\end{table}

\begin{table}[h!]
\begin{center}
\begin{tabular}{cccc}
\hline 
\hline
 & $e_{33}$  & $e_{31}$  & $e_{15}$  \\ 
 ~units~ & (C/cm$^{2}$) & (C/cm$^{2}$) & (C/cm$^{2}$) \\ \hline
~GaN~  	& 0.73 	& -0.49 & -0.40 \\
~InN~	& 0.73	& -0.49	& -0.40 \\
~AlN~  	& 1.55 	& -0.58 & -0.48 \\
\hline
\hline 
\end{tabular} 
\end{center}
\caption{Piezoelectric coefficients for wurtzite III-N materials.\cite{Romanov2006}}
\label{tab:parameter-piezo}
\end{table}

\begin{table}[h!]
\begin{center}
\begin{tabular}{cccc}
\hline 
\hline
 					& $a$	& $b$	& $c$  \\ 
 ~units~ & (C/cm$^{2}$) & (C/cm$^{2}$) & (C/cm$^{2}$) \\ \hline
~In$_x$Ga$_{1-x}$N~	& -0.042	& -0.034	& 0.037 \\
~Al$_x$Ga$_{1-x}$N~	& -0.090	& -0.034	& 0.021 \\
\hline
\hline 
\end{tabular} 
\end{center}
\caption{Parameters of polarization values.\cite{Fiorentini2002, Ambacher2002}}
\label{tab:parameter-polar}
\end{table}

To model the 3D strain distribution in the disordered system, we adopt the 3D continuum strain-stress model solved by the finite element method (FEM) to calculate the strain distribution over the entire LED before solving the Poisson and DD equations.\cite{Hsu2015} The calculated strain is transformed into the piezoelectric polarization field as
\begin{align}
{\bf P}^{pz} = \left[{\bf e}\right] \cdot \left[ {\bf \epsilon} \right] = \left[
\begin{array}{c}
e_{15} \epsilon_{xz}\\ e_{15} \epsilon_{yz}\\e_{31}\left(\epsilon_{xx} + \epsilon_{yy}\right) + e_{33} \epsilon_{zz}
\end{array}\right]
\end{align}
where $\epsilon_{xx}$, $\epsilon_{yy}$, $\epsilon_{zz}$ are normal strains and $\epsilon_{yz}$, $\epsilon_{zx}$, $\epsilon_{xy}$ are shear strains. ${\bf P}^{pz}$ is the strain-induced piezoelectric polarization. $e_{15}$, $e_{31}$, and $e_{33}$ are the piezoelectric coefficients (Table~\ref{tab:parameter-piezo}), and other terms are zero due to the symmetry of wurtzite crystal structures. On the other hand, the spontaneous polarization values related to the GaN buffer layer is obtained by the following equation:
\begin{equation}
{\bf P}^{sp} = a x + b (1-x) + c x(1-x)~,
\end{equation}
where the $a$, $b$, and $c$ coefficients can be found in Table~\ref{tab:parameter-polar}. By taking the divergence of the total polarization ${\bf P}^{\rm total}$, which includes the spontaneous and piezoelectric polarization field (${\bf P}^{\rm total} = {\bf P}^{sp} + {\bf P}^{pz}$) over the entire domain, we calculate $\rho_{pol}(\vec{r})$ as
\begin{equation}
\nabla \cdot {\bf P}^{\rm total} = \rho_{pol}\left({\vec{r}}\right)
\end{equation}
This induced fixed polarization charge $\rho_{pol}$ at different locations is finally implemented into the Poisson equation as the initial condition, as shown in the flowchart of Fig.~\ref{fig:flowchart}.

\section{The localization landscape in nitride LEDs}

In this paper, the size of the simulated domain is 35~nm $\times$ 35~nm $\times$ 387~nm with a full LED structure including six-pairs MQW, an EBL, and $p$ and $n$ transport layers, as shown in Fig.~\ref{fig:schematic_LED}(a). The simulation will model the current injection, transport and carrier screening of polarization fields, by solving the equations self-consistently. The geometric structure was meshed by the Gmsh program,\cite{Geuzaine2009} where the mesh has 1,265,291 nodes and 7,662,428 tetrahedral elements. The mesh grid size is 0.5~nm $\times$ 0.5~nm in the $x$-$y$~plane and a gradual mesh technique was used for the grid size in the $z$-direction ranging from 0.12~nm to 20~nm. The Schr\"odinger equation is of course an eigenvalue problem, whereas the LL~model solves a much simpler linear equation. This significantly reduces the computation time in each iteration step by a factor of $\sim$~1000 as compared to a Schr\"odinger solver. In addition, the computation time of the landscape equation is approximately the same as that for the DD equation (with both electrons and holes). The detailed computation time required to solve each equation and a comparison with other models are given in the Appendix.

\begin{figure}
\includegraphics[width=0.45\textwidth]{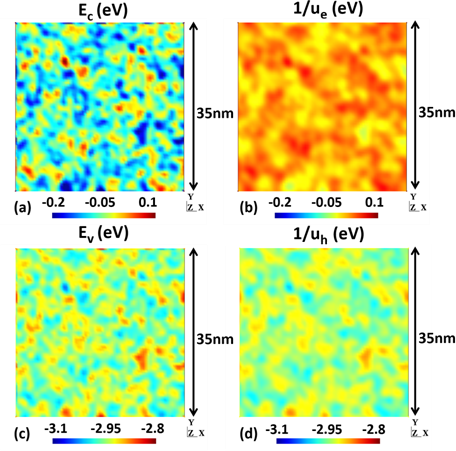}
\caption{In-plane potential energy maps computed in the mid-plane of the third QW of the LED structure at 2.8~V bias. (a) and (c) are the conduction and valence band potentials, respectively, solved by classical Poisson and DD~model. (b) and (d) are the effective confining potentials solved by the LL~theory corresponding to the conduction band and valence band potential, respectively. The location of the plane is displayed in Fig.~\ref{fig:band_diagrams}(b).}
\label{fig:potential_maps}
\end{figure}

\begin{figure}
\includegraphics[width=0.45\textwidth]{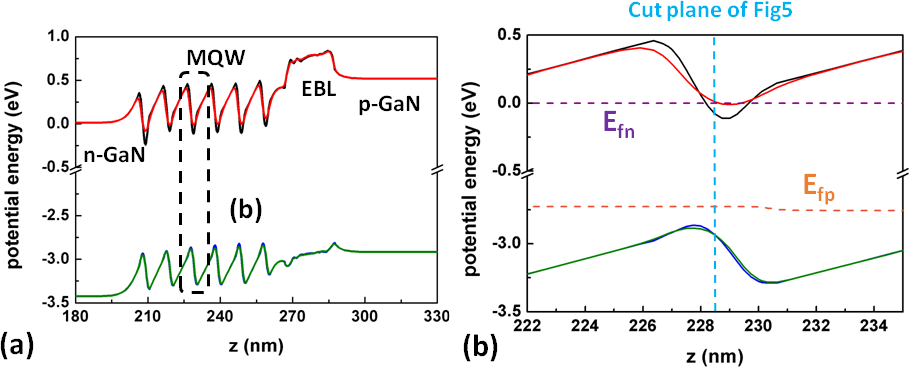}
\caption{(a) Band diagrams of the LED structure at 2.8~V bias along the $z$-direction corresponding to: $E_c$ and $E_v$ (black and blue lines, respectively), $1/u_e$ and $1/u_h$ (red and green lines, respectively) computed self-consistently. (b) Details of the zoomed band diagram, where the variations of the effective potentials $1/u$ with respect to the original band edge potentials can be observed. The electron and hole quasi Fermi levels are shown by dashed lines.}
\label{fig:band_diagrams}
\end{figure}

We now implement the LL~theory into the Poisson-DD model and solve these equations self-consistently (Fig.~\ref{fig:flowchart}) to account for disorders. This solves the carrier density and transport including the quantum effects of disorder inasmuch the landscape model results represent those of a Schr\"odinger solver.\cite{Filoche2017} At a given applied bias to the LED structure the 3D~LL is computed in a self-consistent manner starting from the original electron and hole potentials. As an illustration, the 2D energy potential maps corresponding to the mid-plane of a QW and the 1D band diagram of the structure along the $z$-direction are shown in Fig.~\ref{fig:potential_maps} and Fig.~\ref{fig:band_diagrams}, both for the original and the effective potentials $1/u_{e,h}$ at a bias of 2.8~V. 

Figures~\ref{fig:potential_maps}(a) and \ref{fig:potential_maps}(b) show the conduction band potential and the corresponding landscape potential $1/u_e$ computed self-consistently in presence of QW disorder. $1/u_e$ appears to be smoother compared to $E_c$ because the landscape theory flattens the rapid fluctuations not ``seen'' by the quantum states of the disordered system.\cite{Filoche2017, Arnold2016} The local peak potentials are lowered and smoothed due to quantum tunneling effects. Besides in-plane quantum effects, the energy reference of $1/u_e$ in Fig.~\ref{fig:potential_maps}(b) is also raised with respect to $E_c$ by quantum confinement along $z$ [see Fig.~\ref{fig:band_diagrams}(b)]. Note that the difference between the $E_v$ and $1/u_h$ maps is smaller, as shown in Fig.~\ref{fig:potential_maps}(c) and \ref{fig:potential_maps}(d). This is due to the heavy hole effective mass (1.829~$m_0$), which is much heavier than the electron effective mass (0.159~$m_0$) in the In$_{0.14}$Ga$_{0.86}$N QW. As a result, the quantum confinement and tunneling effects experienced by heavy holes are much reduced.

Similar effects can be clearly observed in the band diagram of the structure shown in Fig.~\ref{fig:band_diagrams} obtained when averaging in-plane $1/u_e$ and $1/u_h$: $1/u_h$ remains fairly close to the valence band edge, while in $1/u_e$ the minima of the conduction band edge are raised considerably and the barriers are appreciably lowered. We emphasize that the LL~theory captures the quantum effects in a disordered layer over a wide range of energies, far from only the ground states, and thus describes the effective band diagram of a quantum semiconductor structure (see LL1, Ref.~\citenum{Filoche2017}), therefore influencing both in-plane and perpendicular carrier transport through the LED. This will prove essential in the threshold voltage for carrier transport through the heterostructure, and is at the root of the proper description of the perpendicular transport, while using layers with homogeneous averaged composition lead to vastly overestimated threshold voltages.

\section{Electrical and optical properties calculated from the localization landscape theory}

After understanding how the LL~theory predicts the effective local potentials for electrons and holes used to calculate the carrier concentrations in a self-consistent loop with the Poisson-DD equations, we move to study current-voltage characteristics, carrier densities and the quantum efficiency of the whole GaN-based LED structure. The simulation parameters can be found in Table~\ref{tab:parameter-simu} including the doping and recombination coefficients. For the sake of comparison we compute different optoelectronic properties of the LED using the landscape-based Poisson-DD solver with random alloy fluctuations (``$1/u$-Poisson-DD'') and compare it to the classical Poisson-DD method with random alloy fluctuations (``Poisson-DD'') or with uniform QWs and EBL (``Poisson-DD (uniform)''), developed in our previous work.\cite{Wu2012} Note that only the first modeling approach takes into account quantum effects due to disorder because of the implementation of the LL~theory.

\begin{figure}
\includegraphics[width=0.45\textwidth]{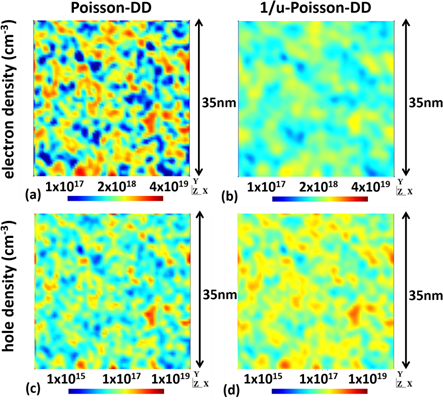}
\caption{Electron and hole carrier densities computed in the mid-plane of the third QW with compositional disorder at 20~A.cm$^{-2}$ current density using: (a),(c) the classical Poisson and drift-diffusion model; (b),(d) the landscape theory implemented in Poisson-DD.}
\label{fig:density_maps}
\end{figure}

In Fig.~\ref{fig:density_maps} we compare the carrier distribution of electrons and holes computed in the mid-plane of the third QW of the LED structure. The classical Poisson and DD solver treats carriers as particles and, as a consequence, the local carrier density fluctuates strongly reflecting the rapid spatial oscillations of the alloy composition [Fig.~\ref{fig:density_maps}(a),(c)].

\begin{table}[tb!]
\begin{center}
\begin{tabular}{lcccc}
\hline 
\hline
& n-GaN	& i-InGaN & p-AlGaN & p-GaN \\
\hline
thickness (nm)	& 200 	& 67 	& 20 	& 100 \\
$\mu_e$ (cm$^2$/Vs)		& 200 	& 300 	& 100 	& 32 \\
$\mu_h$ (cm$^2$/Vs)		& 23	& 10	& 5		& 5 \\
doping (cm$^{-3}$)		& $5\times10^{18}$	& $10^{17}$	& $3\times10^{19}$	& $2\times10^{19}$ \\
$E_a$ (meV)				& 25	& NA	& 215	& 170 \\
$\tau_n^{non rad}$(s)	& 10	& $10^{-7}$	& $10^{-7}$	& $6\times10^{-10}$ \\
$\tau_p^{non rad}$(s)	& $7\times10^{-10}$	& $10^{-7}$	& 10	& 10 \\
$B_0$ (cm$^3$/s)		& $3\times10^{-11}$	& $3\times10^{-11}$	& $3\times10^{-11}$	& $3\times10^{-11}$ \\
$C_0$ (cm$^6$/s)		& $2\times10^{-31}$	& $2\times10^{-31}$	& $2\times10^{-31}$	& $2\times10^{-31}$ \\
\hline
\hline 
\end{tabular} 
\end{center}
\caption{Simulation parameters of each epi-layer.\cite{David2010a, Li2014, Kumakura2005}}
\label{tab:parameter-simu}
\end{table}

On the other hand, in the landscape model quantum effects are included and the smoother effective potentials [Fig.~\ref{fig:potential_maps}(b),(d)], via Eq.~(\ref{eq:carrier_densities}), produce more uniform carrier distributions which are better representative of the standing wave nature of the localized quantum states [Fig.~\ref{fig:density_maps}(b),(d)].

\begin{figure}
\includegraphics[width=0.4\textwidth]{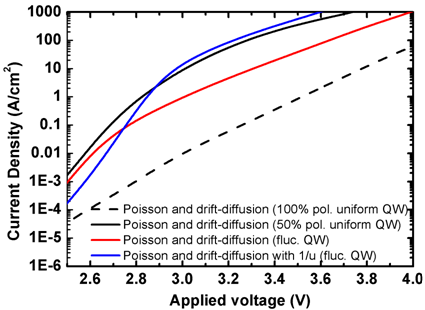}
\caption{A comparison of Poisson DD equations solutions for the I-V characteristics assuming: homogeneous QWs (black curves), disordered QWs described by the real potentials (red curve) or by the $1/u$ effective confining potentials (blue curve).}
\label{fig:I-V}
\end{figure}

The I-V characteristics of the LED calculated by different modeling approaches are shown in Fig.~\ref{fig:I-V}. Usually,\cite{Yang2014} the classical Poisson-DD model without disorder leads to a very large turn-on voltage when assuming 100\% theoretical polarization charge, and in several works~\cite{Piprek2010, Wu2012} an internal charge reduced by 50\% was used to realize a turn-on voltage and IQE more in line with experiment. Both calculations with disorder, ``Poisson-DD'' and ``$1/u$-Poisson-DD'', use the same input random indium distribution and 100\% theoretical polarization charge. Due to the higher effective bandgap of the $1/u$ potentials [Fig.~\ref{fig:band_diagrams}(b)], the current density computed using the landscape is slightly lower before or near turn-on voltage compared to the classical Poisson-DD solver with adjusted 50\% polarization charge. When the applied voltage increases above the threshold, the current density computed using the landscape becomes larger due to the lower effective barriers of the $1/u$ potentials [Fig.~\ref{fig:band_diagrams}(b)], which effectively reduce the internal resistance for carrier transport. The forward voltage ($V_f$) computed by $1/u$ corresponding to 20~A.cm$^{-2}$ is around 3.0~V at 300~K, which matches quite well commercial blue LED data (2.8-3.1~V). For instance, Nichia Co. reported in Ref.~\citenum{Narukawa2010} two blue LEDs, a high-efficiency one with $V_f$=2.89~V at 10~A.cm$^{-2}$ and a high-power one with $V_f$=3.10~V at 35~A.cm$^{-2}$, both values being in good agreement with the I-V characteristic predicted by the $1/u$-Poisson-DD model. The remaining difference could be attributed to leakage paths via V-pit structures or dislocation lines,\cite{Li2016} to the absence in our modeling of tunneling in perpendicular transport, or to the internal temperature of real LED devices being higher than 300~K, leading to a lower $V_f$.

\begin{figure}
\includegraphics[width=0.4\textwidth]{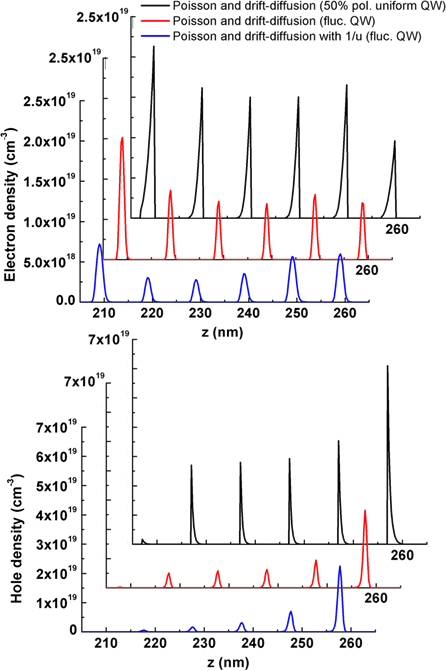}
\caption{Carrier distribution in the 6~QWs of the LED for electrical injection at 20~A.cm$^{-2}$ obtained using a classical Poisson-DD model for uniform layers and the landscape theory implement in Poisson-DD for a structure with random alloy fluctuations. The $x$- and $y$-axis are shifted for illustration.}
\label{fig:carrier_distribution}
\end{figure}

\begin{figure}
\includegraphics[width=0.4\textwidth]{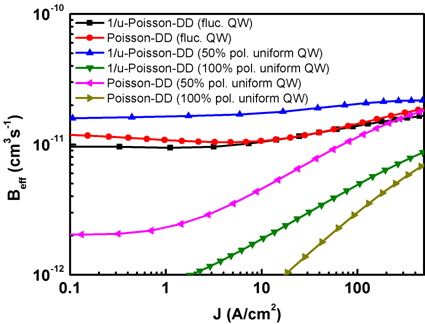}
\caption{Variation of the effective radiative coefficient with carrier injection in the last QW (p-side).}
\label{fig:effective_radiative}
\end{figure}

The average carrier densities along each QW of the active region are shown in Fig.~\ref{fig:carrier_distribution}. The black curves correspond to a structure with uniform In$_{0.14}$Ga$_{0.86}$N QWs and polarization charge reduced to 50\% solved by the classical Poisson and DD model, whereas the blue curves are the QW calculated by the Poisson and drift-diffusion implementing the  landscape.
 
Since at a given LED bias voltage the injected current in the two structures, with and without disorder, can be very different [see Fig.~\ref{fig:I-V}], it is more relevant to compare the carrier densities realized from the different models at a given current density, therefore in comparable conditions of band filling and carrier-induced electric field screening. We also plot in red in Fig.~\ref{fig:carrier_distribution} the calculation of the carrier injection when the disorder is taken into account only through the changes in conduction and valence band levels, without the use of the landscape model to account for localization effects. The results lie somewhat in between those of the uniform material model and the landscape model. As we can see, at the same injected current density of 20~A.cm$^{-2}$, the $1/u$-Poisson-DD model predicts a smaller carrier density as a consequence of the larger electron-hole spatial overlap and higher recombination rates induced by disorder. Moreover the modeling based on the LL shows that carriers are still quite inhomogeneously injected [see Fig.~\ref{fig:QW_distribution}(b)], leaving room to improvement through better designs of the active region as one wishes homogeneous carrier injection to diminish the highly nonlinear Auger recombination.
 
\begin{figure}
\includegraphics[width=0.49\textwidth]{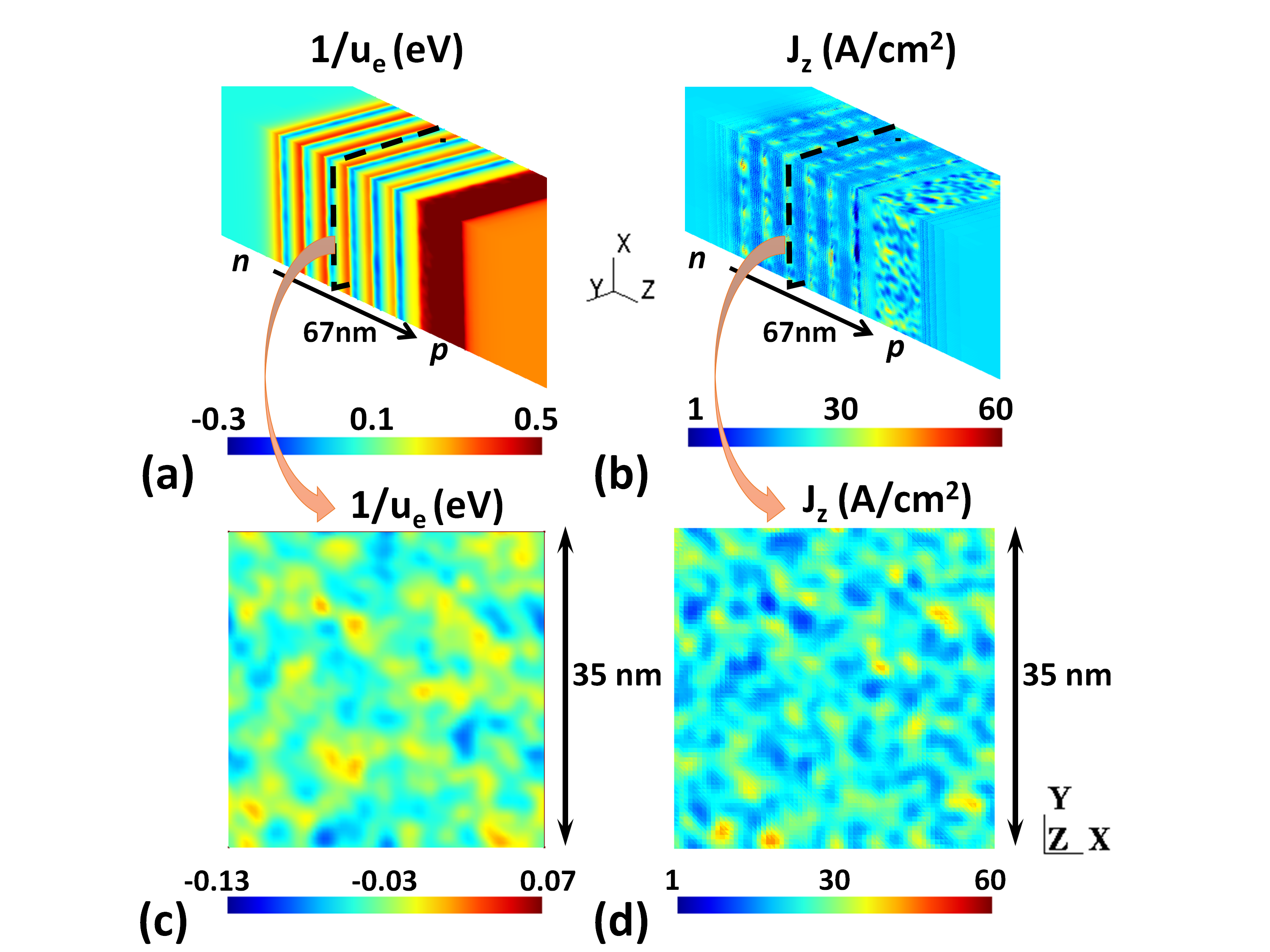}
\caption{(a) and (b) are the perspective views along the $z$-direction of vertical transport of $1/u_e$ and $z$-component of current ($J_z$), respectively. (c) and (d) are the 1/ue and Jz value in the mid-plane ($x$-$y$ plane) of the third QW. All figures are solved by $1/u$-Poisson-DD model, where the LED current density is 20 A.cm$^{-2}$.}
\label{fig:perspective}
\end{figure}

Figure~\ref{fig:effective_radiative} shows the simulated effective $B$ coefficient, $B_{\rm eff}$, for the last QW (p-side) as a function of current density, defined as the recombination rate divided by the product of the QW averaged electron and hole concentrations. As can be seen, $B_{\rm eff}$ increases with current density, mainly due to electric field screening. The increase is particularly important for the simulated uniform QWs, as disorder will smooth out large potential fluctuations. The size of the calculated change in $B_{\rm eff}$ however points that the frequently used ABC model with constant $A$, $B$, $C$ parameters is of little use to quantitatively analyze recombination phenomena in LEDs.\cite{Weisbuch2015}

Figure~\ref{fig:perspective} shows the $z$-component of current and the corresponding effective quantum potentials in the mid-plane ($x$-$y$ plane) of the third QW. As shown in Fig.~\ref{fig:perspective}(d), the current finds percolation paths through lower potential regions, enabling a decrease of the turn-on voltage, where a much smaller voltage is needed to reduce the polarization induced potential barrier. The existence of such percolation paths with high current densities is the basis of the reduced turn-on voltage caused by indium fluctuations induced by disorder, together with the in-plane averaged effects of confinement and tunneling (see Fig.~\ref{fig:band_diagrams}). Similar effects of InGaN alloy fluctuations has been observed and verified in unipolar structures.\cite{Browne2015, Fireman2016}

\begin{figure}
\includegraphics[width=0.35\textwidth]{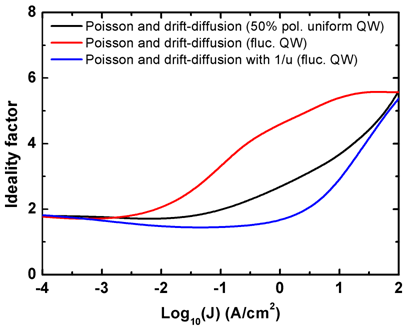}
\caption{Ideality factors corresponding to the I-V characteristics shown in Fig.~\ref{fig:I-V} computed from Eq.~(\ref{eq:IF}).}
\label{fig:ideality}
\end{figure}
 
\section{Discussion}

The above calculation of the I-V characteristic based on $1/u$ only assuming experimental material parameters (Tables I-IV) represents well the experimental data. We turn now to more precise results to identify the calculated LED internal features. Figure~\ref{fig:ideality} displays the ideality factor (IF) of the I-V characteristics calculated as
\begin{equation}\label{eq:IF}
{\rm IF} = \frac{e}{k_B T} \frac{\partial V}{\partial \ln(J)}
\end{equation}
The $1/u$ simulation remarkably reproduces the few experimental data available in high quality LEDs.\cite{AufderMaur2016, Binder2013b, Zhu2009} At low bias and current, IF is near 2, as in the Sah-Noyce-Shockley theory~\cite{Sah1957} due to SRH recombination in the depletion region (here mainly in the QWs). Increasing the bias, the IF diminishes close to unity, as modeled in perfect p-n junctions where current is dominated by diffusion in the neutral regions of the junction. This is to be expected: in the bias region where recombination is dominated by a bimolecular radiative recombination process, the current density is approximately $B_0~np$. Expressing $n$ and $p$ as
\begin{align*}
n &= N_c \exp\left(\frac{E_{Fn}-E_c}{k_B T}\right) \\
p &= N_v \exp\left(\frac{E_v -E_{Fp}}{k_B T}\right)
\end{align*}
This yields
\begin{align*}
J \propto N_c N_v \exp\left(-\frac{E_g}{k_B T}\right) \exp\left(\frac{eV}{k_B T}\right) \approx n_i^2 \exp\left(\frac{eV}{k_B T}\right) ~ ,
\end{align*}
which corresponds to IF=1. The minimum IF, near unity, is reached at a current density of the order of 0.1~A.cm$^{-2}$, in the range of the experimental measurements.\cite{AufderMaur2016, Binder2013b, Zhu2009, David2016} At even larger bias, the IF increases again as the series resistance dominates the device characteristic. Our IF calculation shows that the ``$1/u$-Poisson-DD'' model including random alloy fluctuations provides an excellent overall description of the transport properties of LEDs.

\begin{figure}
\includegraphics[width=0.35\textwidth]{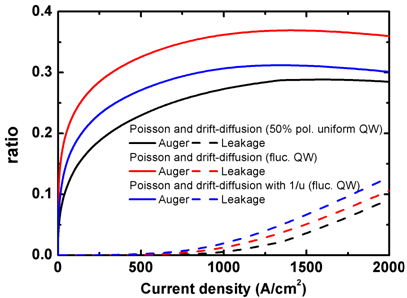}
\caption{Ratio of Auger and leakage currents to total injected current.}
\label{fig:Auger_leakage}
\end{figure}

It must be remarked that an almost ideal IF does not mean at all that carriers are uniformly distributed in the structure (cf. Fig.~\ref{fig:carrier_distribution}) pointing to the inherent shortcomings of electrical measurements to assess that critical phenomenon. We can also calculate the leakage current exiting the active layer region, as shown in Fig.~\ref{fig:Auger_leakage}. As can be expected from the low turn-on voltages modeled, quite smaller than the GaN bandgap, very little carrier leakage is expected under usual operating conditions. The calculation indeed shows that leakage is negligible until the bias voltage reaches 3.4~V, value corresponding to the GaN bandgap, with total currents above 500~A.cm$^{-2}$. In this calculation, the sheet resistance is not considered since only vertical transport is calculated. If sheet resistance would be considered, one would even need higher voltage to reach the flat band condition since an extra potential drop will occur in the current spreading layer.

\begin{figure}
\includegraphics[width=0.35\textwidth]{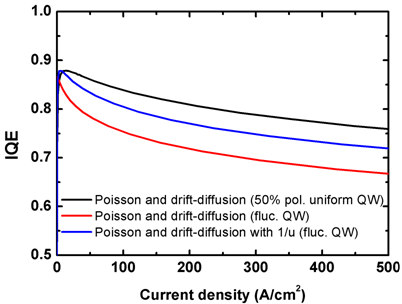}
\caption{IQE curve for the full structure LED.}
\label{fig:IQE}
\end{figure}

\begin{figure}
\includegraphics[width=0.45\textwidth]{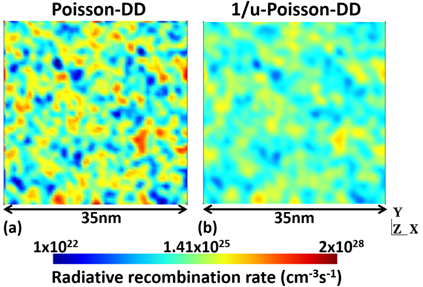}
\caption{The in-plane local radiative recombination rates in the third QW computed from: (a) the classical Poisson-DD model; (b) $1/u$-Poisson-DD model. Current density is fixed at 20~A.cm$^{-2}$.}
\label{fig:radiative_recombination}
\end{figure}

\begin{figure}
\includegraphics[width=0.35\textwidth]{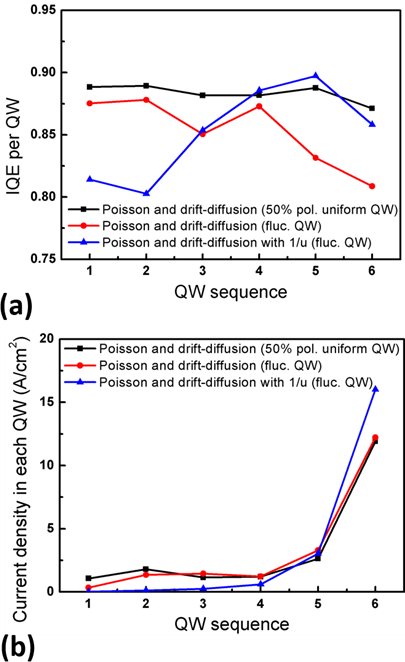}
\caption{Distribution among the different QWs of: (a) IQE, (b) Recombination current density. LED current density is 20~A.cm$^{-2}$.}
\label{fig:QW_distribution}
\end{figure}
  
Turning to the light emission efficiency curve, shown in Fig.~\ref{fig:IQE}, the peak IQE obtained from $1/u$ is slightly higher than the classical Poisson-DD model on fluctuated QWs because of the more homogeneous in-plane radiative efficiency and carrier injection in each QW. As Fig.~\ref{fig:radiative_recombination} shows, the in-plane radiative recombination distribution spreads out in the $1/u$-Poisson-DD model because carriers are localized on larger domains in the QW plane [Fig.~\ref{fig:carrier_distribution}(b),(d)]. Let us recall that the influence of wave function overlap in our model is included in the $np$ term instead of the radiative recombination coefficient $B_0$. Therefore, as shown in Fig.~\ref{fig:carrier_distribution}, the distribution of electrons and holes along the growth direction is more symmetric with respect to the mid-plane of the QWs due to quantum confined wave properties. They have then better overlap compared to the classical DD model, leading to a higher radiative efficiency. 

Figure~\ref{fig:QW_distribution}(b) shows the calculation of the integrated recombined current both radiatively and nonradiatively in each well for the uniform QW and disordered QW models at a current density of 20~A.cm$^{-2}$. The inhomogeneity of carrier injection is well displayed by the total recombination current decreasing from the p-side, on the right, to the n-side. Moreover, the IQE of each QW is also calculated to examine the contribution of each QW as shown in Fig.~\ref{fig:QW_distribution}(a). The IQE of the QW close to the p-layer is smaller in the classical Poisson-DD model due to a smaller overlap as compared to the $1/u$-Poisson-DD model. The opposite trend between the classical Poisson-DD model (higher IQE in the QWs close to the n-layer) and $1/u$-Poisson-DD model (higher IQE in the QWs close to the p-layer) also reveals a better current injection through MQWs for the $1/u$-Poisson-DD model due to reduced effective barriers for electrons. Besides, the Auger recombination will start to dominate in the last QW (p-side well) at higher carrier densities. Due to the increasing local carrier density in the disorder case, the integrated Auger recombination can be enhanced compared to normal QW (see Fig.~\ref{fig:Auger_leakage}), which matches well the experimental droop with a $C_0$ coefficient in line with theory. On the other hand, the different IQE values among QWs can be attributed to the inhomogeneity of carrier injection, which again shows that adopting a constant carrier density ($n$) in the ABC model is incorrect to represent the IQE behavior in MQW structures. While small, the recombination currents in QWs other than the last one cannot be neglected, as is sometimes done in ABC modeling. In the 50\% polarization uniform QWs case, the intentionally reduced polarization field makes electrons and holes overlap better due to the more homogeneous distribution in each well, which does not match experimental observations.\cite{David2008} It also reduces the influence of Auger recombination due to lower local carrier density (without localization), so that the droop effect is less pronounced. A larger $C_0$ would then be needed to represent the experimental Auger recombination.

\section{Conclusions}

In this work, we successfully implemented a novel method, namely the LL~theory of disordered systems, to model the carrier transport and optical emission of LED heterostructures including the effects of intrinsic disorder in nitride-based material alloys. According to the LL~theory, a function $1/u$ acts as an effective semiclassical confining potential which allows us to account for in-plane quantum confinement and tunneling effects due to the random indium fluctuations. The carrier dynamics can then be modeled through the classical drift-diffusion equations in an efficient self-consistent way. With the landscape model, computations are much faster than the conventional Schr\"odinger eigensolvers, especially in 3D, typically by a factor $\times$100-1000, allowing self-consistent calculations.

The I-V characterization of LEDs matches very well experimental measurements as a result of reduced energy barriers and of percolative transport. While the ideality factor of the LED has a near perfect behavior as a function of injected current, the carrier distributions are still very inhomogeneous throughout the heterostructures. This shows that measuring electrically the structure does not provide a clear insight on the internal electronic processes. On the optical properties side, the landscape maps for electrons and holes give us a good estimate of electron and hole overlap, leading to more accurate simulations of LED IQE. 

In principle, this method is not only restricted to modeling nitride-based devices, but can be expanded to model other disordered semiconductor materials and structures.

\begin{acknowledgments}
This work was supported by the Ministry of Science and Technology in Taiwan for Taiwan-ANR collaboration project under Grant MOST 104-2923-E-002-004-MY3, 105-2221-E-002-098-MY3 and by the French National Research Agency (ANR) under grant ANR-14-CE05-0048-01. Svitlana Mayboroda is partially supported by the Alfred P. Sloan Fellowship, the NSF CAREER Award DMS-1056004, the NSF MRSEC Seed Grant, and the NSF INSPIRE Grant. Additional support for Marco Piccardo, James S. Speck and Claude Weisbuch was provided by the DOE Solid State Lighting Program under Award \# DE-EE0007096.
\end{acknowledgments}

\appendix*

\section{Computation time of the landscape theory applied to LEDs compared to other models}

Table V shows the approximate computation time required for each iteration step when solving different equations, such as the Poisson, drift-diffusion, LL and Schr\"odinger equation. The total computation time required to compute a typical LED I-V curve (45~bias values) using different methods is given in Table~\ref{tab:computation_time}, where an average number of 16 iteration steps until convergence is needed for each bias (Poisson-DD and Poisson-DD-$1/u$). The ARPACK solver~\cite{Lehoucq1997} for the generalized eigenvalue problem was used. For the solution of the inverse problem, the PARDISO solver~\cite{Schenk2001} was used. Typically, the environment of clusters to which the computation is submitted is 2 Intel Xeon E5-5650V2 8 cores 2.6~GHz CPUs with 396GB memory.

Let us draw a comparison with the computation time deduced from the simulations of InGaN/GaN QWs incorporating random indium fluctuations as reported by other groups (Table~\ref{tab:iteration_time}). D.~Watson-Parris et al. used the finite difference method to solve the 3-D effective mass Schr\"odinger equation,\cite{WatsonParris2011} where the node size is about 1,500,000. Their computation time for one iteration is about 60,000 seconds, which is quite similar to our Schr\"odinger simulation model and is extremely time consuming. Besides, the self-consistent loop cannot be done due to this long computation time so that the Poisson-Schr\"odinger solver is not self-consistent. Concerning atomistic simulations, S.~Schulz et al. adopted the empirical tight binding method (TBM) and valence force field model to account for the strain-induced polarization field and band structure of the QW, while the perpendicular carrier transport was overlooked.\cite{Schulz2015} The simulation domain is limited near the single QW region of 10~nm $\times$ 9~nm $\times$ 10~nm containing $\sim$~82,000 atoms. Hence, the typical node size is about 82,000$\times$4, with an estimated computation time of 7,500~seconds. M.~Auf~der~Maur et al. also applied the empirical TBM to model indium fluctuations in InGaN QWs.\cite{AufderMaur2016} Although the classical Poisson and drift-diffusion model was used to solve the electrostatic potential, the atomistic calculation is not performed self-consistently with the classical model. The dimension of the atomistic simulation is still limited near the QW region (10~nm $\times$ 10~nm $\times$ 11~nm containing $\sim$~100,000 atoms), which cannot be used to model full MQW LED structure in view of the computation time. Therefore, the key problem for quantum solvers such as the effective mass approximation Schr\"odinger equation or TBM, is the huge amount of computation time which makes such models impractical for the full structure LED simulation.

It can be concluded that the landscape model coupled to the Poisson-DD equations is much more computationally efficient with respect to state-of-the-art quantum solvers, while still incorporating quantum effects such as tunneling and quantum confinement.

\begin{table}
\begin{center}
\begin{tabular}{lcc}
\hline
\hline
							& Node number	& Computation time~(s)	\\
							& (matrix size)	& \\
\hline
Poisson						& 428,655	& 25 \\
Drift-diffusion				& 428,655	& 50 \\
Localization landscape		& 428,655	& 50 \\
Schr\"odinger				& 428,655	& 63,650 \\
Ref.~\citenum{WatsonParris2011}, \citenum{WatsonParris2011b}	& 1,500,000	& 60,000 \\
Ref.~\citenum{Schulz2015}		& 328,000	& 7,500 \\
Ref.~\citenum{AufderMaur2016}	& 100,000	& 24,000 \\
\hline
\hline
\end{tabular}
\end{center}
\caption{Computation time required for each iteration step when solving the Poisson, DD, landscape, and Schr\"odinger equations for a given number of nodes, as tested by our home-built software and compared with other disorder models in nitrides.\cite{WatsonParris2011, Schulz2015, AufderMaur2016, WatsonParris2011b}}
\label{tab:iteration_time}
\end{table}

\begin{table}[ht!]
\begin{center}
\begin{tabular}{lc}
\hline
\hline
							& Total computation time (s) \\
							\hline
Poisson-DD					& 54,000 \\
Poisson-DD-Schr\"odinger	& *45,882,000 \\
Poisson-DD-$1/u$			& 90,000 \\
\hline
\hline
\end{tabular}
\end{center}
\caption{Total computation time of a complete I-V curve ranging from 1.8~V to 4.0~V (0.05~V step) for the Poisson-DD solver, Poisson-DD-Schr\"odinger solver, and Poisson-DD-$1/u$ solver, as tested by our home-built software. (*: This value has been obtained from the Table~\ref{tab:iteration_time} assuming 16 iterations to achieve convergence.)}
\label{tab:computation_time}
\end{table}


\bibliography{localization}

\begin{thebibliography}{62}%
\makeatletter
\providecommand \@ifxundefined [1]{%
 \@ifx{#1\undefined}
}%
\providecommand \@ifnum [1]{%
 \ifnum #1\expandafter \@firstoftwo
 \else \expandafter \@secondoftwo
 \fi
}%
\providecommand \@ifx [1]{%
 \ifx #1\expandafter \@firstoftwo
 \else \expandafter \@secondoftwo
 \fi
}%
\providecommand \natexlab [1]{#1}%
\providecommand \enquote  [1]{``#1''}%
\providecommand \bibnamefont  [1]{#1}%
\providecommand \bibfnamefont [1]{#1}%
\providecommand \citenamefont [1]{#1}%
\providecommand \href@noop [0]{\@secondoftwo}%
\providecommand \href [0]{\begingroup \@sanitize@url \@href}%
\providecommand \@href[1]{\@@startlink{#1}\@@href}%
\providecommand \@@href[1]{\endgroup#1\@@endlink}%
\providecommand \@sanitize@url [0]{\catcode `\\12\catcode `\$12\catcode
  `\&12\catcode `\#12\catcode `\^12\catcode `\_12\catcode `\%12\relax}%
\providecommand \@@startlink[1]{}%
\providecommand \@@endlink[0]{}%
\providecommand \url  [0]{\begingroup\@sanitize@url \@url }%
\providecommand \@url [1]{\endgroup\@href {#1}{\urlprefix }}%
\providecommand \urlprefix  [0]{URL }%
\providecommand \Eprint [0]{\href }%
\providecommand \doibase [0]{http://dx.doi.org/}%
\providecommand \selectlanguage [0]{\@gobble}%
\providecommand \bibinfo  [0]{\@secondoftwo}%
\providecommand \bibfield  [0]{\@secondoftwo}%
\providecommand \translation [1]{[#1]}%
\providecommand \BibitemOpen [0]{}%
\providecommand \bibitemStop [0]{}%
\providecommand \bibitemNoStop [0]{.\EOS\space}%
\providecommand \EOS [0]{\spacefactor3000\relax}%
\providecommand \BibitemShut  [1]{\csname bibitem#1\endcsname}%
\let\auto@bib@innerbib\@empty
%</preamble>
\bibitem [{\citenamefont {Filoche}\ \emph {et~al.}(2017)\citenamefont
  {Filoche}, \citenamefont {Piccardo}, \citenamefont {Weisbuch}, \citenamefont
  {Li}, \citenamefont {Wu},\ and\ \citenamefont {Mayboroda}}]{Filoche2017}%
  \BibitemOpen
  \bibfield  {author} {\bibinfo {author} {\bibfnamefont {M.}~\bibnamefont
  {Filoche}}, \bibinfo {author} {\bibfnamefont {M.}~\bibnamefont {Piccardo}},
  \bibinfo {author} {\bibfnamefont {C.}~\bibnamefont {Weisbuch}}, \bibinfo
  {author} {\bibfnamefont {C.-K.}\ \bibnamefont {Li}}, \bibinfo {author}
  {\bibfnamefont {Y.-R.}\ \bibnamefont {Wu}}, \ and\ \bibinfo {author}
  {\bibfnamefont {S.}~\bibnamefont {Mayboroda}},\ }\href@noop {} {\bibfield
  {journal} {\bibinfo  {journal} {Physical Review B}\ } (\bibinfo {year}
  {2017})}\BibitemShut {NoStop}%
\bibitem [{\citenamefont {Piccardo}\ \emph {et~al.}(2017)\citenamefont
  {Piccardo}, \citenamefont {Li}, \citenamefont {Wu}, \citenamefont {Speck},
  \citenamefont {Bonef}, \citenamefont {Farrell}, \citenamefont {Filoche},
  \citenamefont {Martinelli}, \citenamefont {Peretti},\ and\ \citenamefont
  {Weisbuch}}]{Piccardo2017}%
  \BibitemOpen
  \bibfield  {author} {\bibinfo {author} {\bibfnamefont {M.}~\bibnamefont
  {Piccardo}}, \bibinfo {author} {\bibfnamefont {C.-K.}\ \bibnamefont {Li}},
  \bibinfo {author} {\bibfnamefont {Y.-R.}\ \bibnamefont {Wu}}, \bibinfo
  {author} {\bibfnamefont {J.~S.}\ \bibnamefont {Speck}}, \bibinfo {author}
  {\bibfnamefont {B.}~\bibnamefont {Bonef}}, \bibinfo {author} {\bibfnamefont
  {R.~M.}\ \bibnamefont {Farrell}}, \bibinfo {author} {\bibfnamefont
  {M.}~\bibnamefont {Filoche}}, \bibinfo {author} {\bibfnamefont
  {L.}~\bibnamefont {Martinelli}}, \bibinfo {author} {\bibfnamefont
  {J.}~\bibnamefont {Peretti}}, \ and\ \bibinfo {author} {\bibfnamefont
  {C.}~\bibnamefont {Weisbuch}},\ }\href@noop {} {\bibfield  {journal}
  {\bibinfo  {journal} {Physical Review B}\ } (\bibinfo {year}
  {2017})}\BibitemShut {NoStop}%
\bibitem [{\citenamefont {Weisbuch}\ \emph {et~al.}(2015)\citenamefont
  {Weisbuch}, \citenamefont {Piccardo}, \citenamefont {Martinelli},
  \citenamefont {Iveland}, \citenamefont {Peretti},\ and\ \citenamefont
  {Speck}}]{Weisbuch2015}%
  \BibitemOpen
  \bibfield  {author} {\bibinfo {author} {\bibfnamefont {C.}~\bibnamefont
  {Weisbuch}}, \bibinfo {author} {\bibfnamefont {M.}~\bibnamefont {Piccardo}},
  \bibinfo {author} {\bibfnamefont {L.}~\bibnamefont {Martinelli}}, \bibinfo
  {author} {\bibfnamefont {J.}~\bibnamefont {Iveland}}, \bibinfo {author}
  {\bibfnamefont {J.}~\bibnamefont {Peretti}}, \ and\ \bibinfo {author}
  {\bibfnamefont {J.~S.}\ \bibnamefont {Speck}},\ }\href@noop {} {\bibfield
  {journal} {\bibinfo  {journal} {Physica Status Solidi (a)}\ }\textbf
  {\bibinfo {volume} {212}},\ \bibinfo {pages} {899} (\bibinfo {year}
  {2015})}\BibitemShut {NoStop}%
\bibitem [{\citenamefont {Nakamura}\ \emph {et~al.}(1991)\citenamefont
  {Nakamura}, \citenamefont {Mukai},\ and\ \citenamefont
  {Senoh}}]{Nakamura1991}%
  \BibitemOpen
  \bibfield  {author} {\bibinfo {author} {\bibfnamefont {S.}~\bibnamefont
  {Nakamura}}, \bibinfo {author} {\bibfnamefont {T.}~\bibnamefont {Mukai}}, \
  and\ \bibinfo {author} {\bibfnamefont {M.}~\bibnamefont {Senoh}},\
  }\href@noop {} {\bibfield  {journal} {\bibinfo  {journal} {Japanese Journal
  of Applied Physics}\ }\textbf {\bibinfo {volume} {30}},\ \bibinfo {pages}
  {L1998} (\bibinfo {year} {1991})}\BibitemShut {NoStop}%
\bibitem [{\citenamefont {Pengelly}\ \emph {et~al.}(2012)\citenamefont
  {Pengelly}, \citenamefont {Wood}, \citenamefont {Milligan}, \citenamefont
  {Sheppard},\ and\ \citenamefont {Pribble}}]{Pengelly2012}%
  \BibitemOpen
  \bibfield  {author} {\bibinfo {author} {\bibfnamefont {R.~S.}\ \bibnamefont
  {Pengelly}}, \bibinfo {author} {\bibfnamefont {S.~M.}\ \bibnamefont {Wood}},
  \bibinfo {author} {\bibfnamefont {J.~W.}\ \bibnamefont {Milligan}}, \bibinfo
  {author} {\bibfnamefont {S.~T.}\ \bibnamefont {Sheppard}}, \ and\ \bibinfo
  {author} {\bibfnamefont {W.~L.}\ \bibnamefont {Pribble}},\ }\href {\doibase
  10.1109/TMTT.2012.2187535} {\bibfield  {journal} {\bibinfo  {journal} {IEEE
  Transactions on Microwave Theory and Techniques}\ }\textbf {\bibinfo {volume}
  {60}},\ \bibinfo {pages} {1764} (\bibinfo {year} {2012})}\BibitemShut
  {NoStop}%
\bibitem [{\citenamefont {Kim}\ \emph {et~al.}(2007)\citenamefont {Kim},
  \citenamefont {Schubert}, \citenamefont {Dai}, \citenamefont {Kim},
  \citenamefont {Schubert}, \citenamefont {Piprek}, ,\ and\ \citenamefont
  {Park}}]{Kim2007}%
  \BibitemOpen
  \bibfield  {author} {\bibinfo {author} {\bibfnamefont {M.-H.}\ \bibnamefont
  {Kim}}, \bibinfo {author} {\bibfnamefont {M.~F.}\ \bibnamefont {Schubert}},
  \bibinfo {author} {\bibfnamefont {Q.}~\bibnamefont {Dai}}, \bibinfo {author}
  {\bibfnamefont {J.~K.}\ \bibnamefont {Kim}}, \bibinfo {author} {\bibfnamefont
  {E.~F.}\ \bibnamefont {Schubert}}, \bibinfo {author} {\bibfnamefont
  {J.}~\bibnamefont {Piprek}}, , \ and\ \bibinfo {author} {\bibfnamefont
  {Y.}~\bibnamefont {Park}},\ }\href@noop {} {\bibfield  {journal} {\bibinfo
  {journal} {Applied Physics Letters}\ }\textbf {\bibinfo {volume} {91}},\
  \bibinfo {pages} {183507} (\bibinfo {year} {2007})}\BibitemShut {NoStop}%
\bibitem [{\citenamefont {\"Ozg\"ur}\ \emph {et~al.}(2011)\citenamefont
  {\"Ozg\"ur}, \citenamefont {Ni}, \citenamefont {Li}, \citenamefont {Lee},
  \citenamefont {Liu}, \citenamefont {Okur}, \citenamefont {Avrutin},
  \citenamefont {Matulionis},\ and\ \citenamefont {Morko\c{c}}}]{Ozgur2011}%
  \BibitemOpen
  \bibfield  {author} {\bibinfo {author} {\bibfnamefont {U.}~\bibnamefont
  {\"Ozg\"ur}}, \bibinfo {author} {\bibfnamefont {X.}~\bibnamefont {Ni}},
  \bibinfo {author} {\bibfnamefont {X.}~\bibnamefont {Li}}, \bibinfo {author}
  {\bibfnamefont {J.}~\bibnamefont {Lee}}, \bibinfo {author} {\bibfnamefont
  {S.}~\bibnamefont {Liu}}, \bibinfo {author} {\bibfnamefont {S.}~\bibnamefont
  {Okur}}, \bibinfo {author} {\bibfnamefont {V.}~\bibnamefont {Avrutin}},
  \bibinfo {author} {\bibfnamefont {A.}~\bibnamefont {Matulionis}}, \ and\
  \bibinfo {author} {\bibfnamefont {H.}~\bibnamefont {Morko\c{c}}},\ }\href
  {\doibase 10.1088/0268-1242/26/1/014022} {\bibfield  {journal} {\bibinfo
  {journal} {Semiconductor Science and Technology}\ }\textbf {\bibinfo {volume}
  {26}},\ \bibinfo {pages} {014022} (\bibinfo {year} {2011})}\BibitemShut
  {NoStop}%
\bibitem [{\citenamefont {Iveland}\ \emph {et~al.}(2013)\citenamefont
  {Iveland}, \citenamefont {Martinelli}, \citenamefont {Peretti}, \citenamefont
  {Speck},\ and\ \citenamefont {Weisbuch}}]{Iveland2013}%
  \BibitemOpen
  \bibfield  {author} {\bibinfo {author} {\bibfnamefont {J.}~\bibnamefont
  {Iveland}}, \bibinfo {author} {\bibfnamefont {L.}~\bibnamefont {Martinelli}},
  \bibinfo {author} {\bibfnamefont {J.}~\bibnamefont {Peretti}}, \bibinfo
  {author} {\bibfnamefont {J.~S.}\ \bibnamefont {Speck}}, \ and\ \bibinfo
  {author} {\bibfnamefont {C.}~\bibnamefont {Weisbuch}},\ }\href@noop {}
  {\bibfield  {journal} {\bibinfo  {journal} {Physical Review Letters}\
  }\textbf {\bibinfo {volume} {110}},\ \bibinfo {pages} {177406} (\bibinfo
  {year} {2013})}\BibitemShut {NoStop}%
\bibitem [{\citenamefont {Shen}\ \emph {et~al.}(2007)\citenamefont {Shen},
  \citenamefont {Mueller}, \citenamefont {Watanabe}, \citenamefont {Gardner},
  \citenamefont {Munkholm},\ and\ \citenamefont {Krames}}]{Shen2007}%
  \BibitemOpen
  \bibfield  {author} {\bibinfo {author} {\bibfnamefont {Y.~C.}\ \bibnamefont
  {Shen}}, \bibinfo {author} {\bibfnamefont {G.~O.}\ \bibnamefont {Mueller}},
  \bibinfo {author} {\bibfnamefont {S.}~\bibnamefont {Watanabe}}, \bibinfo
  {author} {\bibfnamefont {N.~F.}\ \bibnamefont {Gardner}}, \bibinfo {author}
  {\bibfnamefont {A.}~\bibnamefont {Munkholm}}, \ and\ \bibinfo {author}
  {\bibfnamefont {M.~R.}\ \bibnamefont {Krames}},\ }\href {\doibase
  10.1063/1.2785135} {\bibfield  {journal} {\bibinfo  {journal} {Applied
  Physics Letters}\ }\textbf {\bibinfo {volume} {91}},\ \bibinfo {pages}
  {141101} (\bibinfo {year} {2007})}\BibitemShut {NoStop}%
\bibitem [{\citenamefont {Binder}\ \emph
  {et~al.}(2013{\natexlab{a}})\citenamefont {Binder}, \citenamefont {Nirschl},
  \citenamefont {Zeisel}, \citenamefont {Hager}, \citenamefont {Lugauer},
  \citenamefont {Sabathil}, \citenamefont {Bougeard}, \citenamefont {Wagner},\
  and\ \citenamefont {Galler}}]{Binder2013a}%
  \BibitemOpen
  \bibfield  {author} {\bibinfo {author} {\bibfnamefont {M.}~\bibnamefont
  {Binder}}, \bibinfo {author} {\bibfnamefont {A.}~\bibnamefont {Nirschl}},
  \bibinfo {author} {\bibfnamefont {R.}~\bibnamefont {Zeisel}}, \bibinfo
  {author} {\bibfnamefont {T.}~\bibnamefont {Hager}}, \bibinfo {author}
  {\bibfnamefont {H.-J.}\ \bibnamefont {Lugauer}}, \bibinfo {author}
  {\bibfnamefont {M.}~\bibnamefont {Sabathil}}, \bibinfo {author}
  {\bibfnamefont {D.}~\bibnamefont {Bougeard}}, \bibinfo {author}
  {\bibfnamefont {J.}~\bibnamefont {Wagner}}, \ and\ \bibinfo {author}
  {\bibfnamefont {B.}~\bibnamefont {Galler}},\ }\href {\doibase
  10.1063/1.4818761} {\bibfield  {journal} {\bibinfo  {journal} {Applied
  Physics Letters}\ }\textbf {\bibinfo {volume} {103}},\ \bibinfo {pages}
  {071108} (\bibinfo {year} {2013}{\natexlab{a}})}\BibitemShut {NoStop}%
\bibitem [{\citenamefont {David}\ and\ \citenamefont
  {Gardner}(2010)}]{David2010b}%
  \BibitemOpen
  \bibfield  {author} {\bibinfo {author} {\bibfnamefont {A.}~\bibnamefont
  {David}}\ and\ \bibinfo {author} {\bibfnamefont {N.~F.}\ \bibnamefont
  {Gardner}},\ }\href {\doibase 10.1063/1.3515851} {\bibfield  {journal}
  {\bibinfo  {journal} {Applied Physics Letters}\ }\textbf {\bibinfo {volume}
  {97}},\ \bibinfo {pages} {193508} (\bibinfo {year} {2010})}\BibitemShut
  {NoStop}%
\bibitem [{\citenamefont {Ni}\ \emph {et~al.}(2008)\citenamefont {Ni},
  \citenamefont {Fan}, \citenamefont {Shimada}, \citenamefont {\"Ozg\"ur},\
  and\ \citenamefont {Morko\c{c}}}]{Ni2008}%
  \BibitemOpen
  \bibfield  {author} {\bibinfo {author} {\bibfnamefont {X.}~\bibnamefont
  {Ni}}, \bibinfo {author} {\bibfnamefont {Q.}~\bibnamefont {Fan}}, \bibinfo
  {author} {\bibfnamefont {R.}~\bibnamefont {Shimada}}, \bibinfo {author}
  {\bibfnamefont {U.}~\bibnamefont {\"Ozg\"ur}}, \ and\ \bibinfo {author}
  {\bibfnamefont {H.}~\bibnamefont {Morko\c{c}}},\ }\href@noop {} {\bibfield
  {journal} {\bibinfo  {journal} {Applied Physics Letters}\ }\textbf {\bibinfo
  {volume} {93}},\ \bibinfo {pages} {171113} (\bibinfo {year}
  {2008})}\BibitemShut {NoStop}%
\bibitem [{\citenamefont {Mukai}\ \emph {et~al.}(1999)\citenamefont {Mukai},
  \citenamefont {Yamada},\ and\ \citenamefont {Nakamura}}]{Mukai1999}%
  \BibitemOpen
  \bibfield  {author} {\bibinfo {author} {\bibfnamefont {T.}~\bibnamefont
  {Mukai}}, \bibinfo {author} {\bibfnamefont {M.}~\bibnamefont {Yamada}}, \
  and\ \bibinfo {author} {\bibfnamefont {S.}~\bibnamefont {Nakamura}},\
  }\href@noop {} {\bibfield  {journal} {\bibinfo  {journal} {Japanese Journal
  of Applied Physics}\ }\textbf {\bibinfo {volume} {38}},\ \bibinfo {pages}
  {3976} (\bibinfo {year} {1999})}\BibitemShut {NoStop}%
\bibitem [{\citenamefont {Watson-Parris}\ \emph {et~al.}(2011)\citenamefont
  {Watson-Parris}, \citenamefont {Godfrey}, \citenamefont {Dawson},
  \citenamefont {Oliver}, \citenamefont {Galtrey}, \citenamefont {Kappers},\
  and\ \citenamefont {Humphreys}}]{WatsonParris2011}%
  \BibitemOpen
  \bibfield  {author} {\bibinfo {author} {\bibfnamefont {D.}~\bibnamefont
  {Watson-Parris}}, \bibinfo {author} {\bibfnamefont {M.~J.}\ \bibnamefont
  {Godfrey}}, \bibinfo {author} {\bibfnamefont {P.}~\bibnamefont {Dawson}},
  \bibinfo {author} {\bibfnamefont {R.~A.}\ \bibnamefont {Oliver}}, \bibinfo
  {author} {\bibfnamefont {M.~J.}\ \bibnamefont {Galtrey}}, \bibinfo {author}
  {\bibfnamefont {M.~J.}\ \bibnamefont {Kappers}}, \ and\ \bibinfo {author}
  {\bibfnamefont {C.~J.}\ \bibnamefont {Humphreys}},\ }\href@noop {} {\bibfield
   {journal} {\bibinfo  {journal} {Physical Review B}\ }\textbf {\bibinfo
  {volume} {83}},\ \bibinfo {pages} {115321} (\bibinfo {year}
  {2011})}\BibitemShut {NoStop}%
\bibitem [{\citenamefont {Kioupakis}\ \emph {et~al.}(2015)\citenamefont
  {Kioupakis}, \citenamefont {Rinke}, \citenamefont {Delaney},\ and\
  \citenamefont {{Van De Walle}}}]{Kioupakis2015}%
  \BibitemOpen
  \bibfield  {author} {\bibinfo {author} {\bibfnamefont {E.}~\bibnamefont
  {Kioupakis}}, \bibinfo {author} {\bibfnamefont {P.}~\bibnamefont {Rinke}},
  \bibinfo {author} {\bibfnamefont {K.~T.}\ \bibnamefont {Delaney}}, \ and\
  \bibinfo {author} {\bibfnamefont {C.~G.}\ \bibnamefont {{Van De Walle}}},\
  }\href@noop {} {\bibfield  {journal} {\bibinfo  {journal} {Physical Review
  B}\ }\textbf {\bibinfo {volume} {92}},\ \bibinfo {pages} {035207} (\bibinfo
  {year} {2015})}\BibitemShut {NoStop}%
\bibitem [{\citenamefont {Bennett}\ \emph {et~al.}(2011)\citenamefont
  {Bennett}, \citenamefont {Saxey}, \citenamefont {Kappers}, \citenamefont
  {Barnard}, \citenamefont {Humphreys}, \citenamefont {Smith},\ and\
  \citenamefont {Oliver}}]{Bennett2011}%
  \BibitemOpen
  \bibfield  {author} {\bibinfo {author} {\bibfnamefont {S.~E.}\ \bibnamefont
  {Bennett}}, \bibinfo {author} {\bibfnamefont {D.~W.}\ \bibnamefont {Saxey}},
  \bibinfo {author} {\bibfnamefont {M.~J.}\ \bibnamefont {Kappers}}, \bibinfo
  {author} {\bibfnamefont {J.~S.}\ \bibnamefont {Barnard}}, \bibinfo {author}
  {\bibfnamefont {C.~J.}\ \bibnamefont {Humphreys}}, \bibinfo {author}
  {\bibfnamefont {G.~D.}\ \bibnamefont {Smith}}, \ and\ \bibinfo {author}
  {\bibfnamefont {R.~A.}\ \bibnamefont {Oliver}},\ }\href@noop {} {\bibfield
  {journal} {\bibinfo  {journal} {Applied Physics Letters}\ }\textbf {\bibinfo
  {volume} {99}},\ \bibinfo {pages} {021906} (\bibinfo {year}
  {2011})}\BibitemShut {NoStop}%
\bibitem [{\citenamefont {Shivaraman}\ \emph {et~al.}(2013)\citenamefont
  {Shivaraman}, \citenamefont {Kawaguchi}, \citenamefont {Tanaka},
  \citenamefont {DenBaars}, \citenamefont {Nakamura},\ and\ \citenamefont
  {Speck}}]{Shivaraman2013}%
  \BibitemOpen
  \bibfield  {author} {\bibinfo {author} {\bibfnamefont {R.}~\bibnamefont
  {Shivaraman}}, \bibinfo {author} {\bibfnamefont {Y.}~\bibnamefont
  {Kawaguchi}}, \bibinfo {author} {\bibfnamefont {S.}~\bibnamefont {Tanaka}},
  \bibinfo {author} {\bibfnamefont {S.~P.}\ \bibnamefont {DenBaars}}, \bibinfo
  {author} {\bibfnamefont {S.}~\bibnamefont {Nakamura}}, \ and\ \bibinfo
  {author} {\bibfnamefont {J.~S.}\ \bibnamefont {Speck}},\ }\href {\doibase
  10.1063/1.4812363} {\bibfield  {journal} {\bibinfo  {journal} {Applied
  Physics Letters}\ }\textbf {\bibinfo {volume} {102}},\ \bibinfo {pages}
  {251104} (\bibinfo {year} {2013})}\BibitemShut {NoStop}%
\bibitem [{\citenamefont {Mazumder}\ \emph {et~al.}(2013)\citenamefont
  {Mazumder}, \citenamefont {Esposto}, \citenamefont {Hung}, \citenamefont
  {Mates}, \citenamefont {Rajan},\ and\ \citenamefont {Speck}}]{Mazumder2013}%
  \BibitemOpen
  \bibfield  {author} {\bibinfo {author} {\bibfnamefont {B.}~\bibnamefont
  {Mazumder}}, \bibinfo {author} {\bibfnamefont {M.}~\bibnamefont {Esposto}},
  \bibinfo {author} {\bibfnamefont {T.~H.}\ \bibnamefont {Hung}}, \bibinfo
  {author} {\bibfnamefont {T.}~\bibnamefont {Mates}}, \bibinfo {author}
  {\bibfnamefont {S.}~\bibnamefont {Rajan}}, \ and\ \bibinfo {author}
  {\bibfnamefont {J.~S.}\ \bibnamefont {Speck}},\ }\href@noop {} {\bibfield
  {journal} {\bibinfo  {journal} {Applied Physics Letters}\ }\textbf {\bibinfo
  {volume} {103}},\ \bibinfo {pages} {151601} (\bibinfo {year}
  {2013})}\BibitemShut {NoStop}%
\bibitem [{\citenamefont {Riley}\ \emph {et~al.}(2014)\citenamefont {Riley},
  \citenamefont {Detchprohm}, \citenamefont {Wetzel},\ and\ \citenamefont
  {Lauhon}}]{Riley2014}%
  \BibitemOpen
  \bibfield  {author} {\bibinfo {author} {\bibfnamefont {J.~R.}\ \bibnamefont
  {Riley}}, \bibinfo {author} {\bibfnamefont {T.}~\bibnamefont {Detchprohm}},
  \bibinfo {author} {\bibfnamefont {C.}~\bibnamefont {Wetzel}}, \ and\ \bibinfo
  {author} {\bibfnamefont {L.~J.}\ \bibnamefont {Lauhon}},\ }\href {\doibase
  10.1063/1.4871510} {\bibfield  {journal} {\bibinfo  {journal} {Applied
  Physics Letters}\ }\textbf {\bibinfo {volume} {104}},\ \bibinfo {pages}
  {152102} (\bibinfo {year} {2014})}\BibitemShut {NoStop}%
\bibitem [{\citenamefont {Schulz}\ \emph {et~al.}(2015)\citenamefont {Schulz},
  \citenamefont {Caro}, \citenamefont {Coughlan},\ and\ \citenamefont
  {O'Reilly}}]{Schulz2015}%
  \BibitemOpen
  \bibfield  {author} {\bibinfo {author} {\bibfnamefont {S.}~\bibnamefont
  {Schulz}}, \bibinfo {author} {\bibfnamefont {M.~A.}\ \bibnamefont {Caro}},
  \bibinfo {author} {\bibfnamefont {C.}~\bibnamefont {Coughlan}}, \ and\
  \bibinfo {author} {\bibfnamefont {E.~P.}\ \bibnamefont {O'Reilly}},\ }\href
  {\doibase 10.1103/PhysRevB.91.035439} {\bibfield  {journal} {\bibinfo
  {journal} {Physical Review B}\ }\textbf {\bibinfo {volume} {91}},\ \bibinfo
  {pages} {035439} (\bibinfo {year} {2015})}\BibitemShut {NoStop}%
\bibitem [{\citenamefont {{Auf der Maur}}\ \emph {et~al.}(2016)\citenamefont
  {{Auf der Maur}}, \citenamefont {Pecchia}, \citenamefont {Penazzi},
  \citenamefont {Rodrigues},\ and\ \citenamefont {{Di
  Carlo}}}]{AufderMaur2016}%
  \BibitemOpen
  \bibfield  {author} {\bibinfo {author} {\bibfnamefont {M.}~\bibnamefont {{Auf
  der Maur}}}, \bibinfo {author} {\bibfnamefont {A.}~\bibnamefont {Pecchia}},
  \bibinfo {author} {\bibfnamefont {G.}~\bibnamefont {Penazzi}}, \bibinfo
  {author} {\bibfnamefont {W.}~\bibnamefont {Rodrigues}}, \ and\ \bibinfo
  {author} {\bibfnamefont {A.}~\bibnamefont {{Di Carlo}}},\ }\href {\doibase
  10.1103/PhysRevLett.116.027401} {\bibfield  {journal} {\bibinfo  {journal}
  {Physical Review Letters}\ }\textbf {\bibinfo {volume} {116}},\ \bibinfo
  {pages} {027401} (\bibinfo {year} {2016})}\BibitemShut {NoStop}%
\bibitem [{\citenamefont {Yang}\ \emph {et~al.}(2014)\citenamefont {Yang},
  \citenamefont {Shivaraman}, \citenamefont {Speck},\ and\ \citenamefont
  {Wu}}]{Yang2014}%
  \BibitemOpen
  \bibfield  {author} {\bibinfo {author} {\bibfnamefont {T.-J.}\ \bibnamefont
  {Yang}}, \bibinfo {author} {\bibfnamefont {R.}~\bibnamefont {Shivaraman}},
  \bibinfo {author} {\bibfnamefont {J.~S.}\ \bibnamefont {Speck}}, \ and\
  \bibinfo {author} {\bibfnamefont {Y.-R.}\ \bibnamefont {Wu}},\ }\href
  {\doibase 10.1063/1.4896103} {\bibfield  {journal} {\bibinfo  {journal}
  {Journal of Applied Physics}\ }\textbf {\bibinfo {volume} {116}},\ \bibinfo
  {pages} {113104} (\bibinfo {year} {2014})}\BibitemShut {NoStop}%
\bibitem [{\citenamefont {Karpov}(2011)}]{Karpov2011}%
  \BibitemOpen
  \bibfield  {author} {\bibinfo {author} {\bibfnamefont {S.~Y.}\ \bibnamefont
  {Karpov}},\ }\href {\doibase 10.1117/12.872842} {\bibfield  {journal}
  {\bibinfo  {journal} {SPIE Proceedings}\ }\textbf {\bibinfo {volume} {7939}}
  (\bibinfo {year} {2011}),\ 10.1117/12.872842}\BibitemShut {NoStop}%
\bibitem [{\citenamefont {Chen}\ \emph {et~al.}(2009)\citenamefont {Chen},
  \citenamefont {Ling}, \citenamefont {Huang}, \citenamefont {Su},
  \citenamefont {Ko}, \citenamefont {Lu}, \citenamefont {Kuo}, \citenamefont
  {Kuo},\ and\ \citenamefont {Wang}}]{Chen2009}%
  \BibitemOpen
  \bibfield  {author} {\bibinfo {author} {\bibfnamefont {J.-R.}\ \bibnamefont
  {Chen}}, \bibinfo {author} {\bibfnamefont {S.-C.}\ \bibnamefont {Ling}},
  \bibinfo {author} {\bibfnamefont {H.-M.}\ \bibnamefont {Huang}}, \bibinfo
  {author} {\bibfnamefont {P.-Y.}\ \bibnamefont {Su}}, \bibinfo {author}
  {\bibfnamefont {T.-S.}\ \bibnamefont {Ko}}, \bibinfo {author} {\bibfnamefont
  {T.-C.}\ \bibnamefont {Lu}}, \bibinfo {author} {\bibfnamefont {H.-C.}\
  \bibnamefont {Kuo}}, \bibinfo {author} {\bibfnamefont {Y.-K.}\ \bibnamefont
  {Kuo}}, \ and\ \bibinfo {author} {\bibfnamefont {S.-C.}\ \bibnamefont
  {Wang}},\ }\href {\doibase 10.1007/s00340-008-3331-9} {\bibfield  {journal}
  {\bibinfo  {journal} {Applied Physics B}\ }\textbf {\bibinfo {volume} {95}},\
  \bibinfo {pages} {145 } (\bibinfo {year} {2009})}\BibitemShut {NoStop}%
\bibitem [{\citenamefont {Browne}\ \emph {et~al.}(2015)\citenamefont {Browne},
  \citenamefont {Mazumder}, \citenamefont {Wu},\ and\ \citenamefont
  {Speck}}]{Browne2015}%
  \BibitemOpen
  \bibfield  {author} {\bibinfo {author} {\bibfnamefont {D.~A.}\ \bibnamefont
  {Browne}}, \bibinfo {author} {\bibfnamefont {B.}~\bibnamefont {Mazumder}},
  \bibinfo {author} {\bibfnamefont {Y.-R.}\ \bibnamefont {Wu}}, \ and\ \bibinfo
  {author} {\bibfnamefont {J.~S.}\ \bibnamefont {Speck}},\ }\href {\doibase
  10.1063/1.4919750} {\bibfield  {journal} {\bibinfo  {journal} {Journal of
  Applied Physics}\ }\textbf {\bibinfo {volume} {117}},\ \bibinfo {pages}
  {185703} (\bibinfo {year} {2015})}\BibitemShut {NoStop}%
\bibitem [{\citenamefont {Piprek}(2010)}]{Piprek2010}%
  \BibitemOpen
  \bibfield  {author} {\bibinfo {author} {\bibfnamefont {J.}~\bibnamefont
  {Piprek}},\ }\href {\doibase 10.1002/pssa.201026149} {\bibfield  {journal}
  {\bibinfo  {journal} {Physica Status Solidi (a)}\ }\textbf {\bibinfo {volume}
  {207}},\ \bibinfo {pages} {2217} (\bibinfo {year} {2010})}\BibitemShut
  {NoStop}%
\bibitem [{\citenamefont {Wu}\ \emph {et~al.}(2012)\citenamefont {Wu},
  \citenamefont {Shivaraman}, \citenamefont {Wang},\ and\ \citenamefont
  {Speck}}]{Wu2012}%
  \BibitemOpen
  \bibfield  {author} {\bibinfo {author} {\bibfnamefont {Y.-R.}\ \bibnamefont
  {Wu}}, \bibinfo {author} {\bibfnamefont {R.}~\bibnamefont {Shivaraman}},
  \bibinfo {author} {\bibfnamefont {K.-C.}\ \bibnamefont {Wang}}, \ and\
  \bibinfo {author} {\bibfnamefont {J.~S.}\ \bibnamefont {Speck}},\ }\href
  {http://scitation.aip.org/content/aip/journal/apl/101/8/10.1063/1.4747532}
  {\bibfield  {journal} {\bibinfo  {journal} {Applied Physics Letters}\
  }\textbf {\bibinfo {volume} {101}},\ \bibinfo {eid} {083505} (\bibinfo {year}
  {2012})}\BibitemShut {NoStop}%
\bibitem [{\citenamefont {Shedbalkar}\ \emph {et~al.}(2016)\citenamefont
  {Shedbalkar}, \citenamefont {Andreev},\ and\ \citenamefont
  {Witzigmann}}]{Shedbalkar2016}%
  \BibitemOpen
  \bibfield  {author} {\bibinfo {author} {\bibfnamefont {A.}~\bibnamefont
  {Shedbalkar}}, \bibinfo {author} {\bibfnamefont {Z.}~\bibnamefont {Andreev}},
  \ and\ \bibinfo {author} {\bibfnamefont {B.}~\bibnamefont {Witzigmann}},\
  }\href {\doibase 10.1002/pssb.201552276} {\bibfield  {journal} {\bibinfo
  {journal} {Physica Status Solidi (b)}\ }\textbf {\bibinfo {volume} {253}},\
  \bibinfo {pages} {158} (\bibinfo {year} {2016})}\BibitemShut {NoStop}%
\bibitem [{\citenamefont {Datta}(1995)}]{Datta1995}%
  \BibitemOpen
  \bibfield  {author} {\bibinfo {author} {\bibfnamefont {S.}~\bibnamefont
  {Datta}},\ }\href@noop {} {\emph {\bibinfo {title} {Electronic Transport in
  Mesoscopic Systems}}},\ Cambridge Studies in Semiconductor Physics and
  Microelectronic Engineering\ (\bibinfo  {publisher} {Cambridge University
  Press},\ \bibinfo {year} {1995})\BibitemShut {NoStop}%
\bibitem [{\citenamefont {Wu}\ \emph {et~al.}(2015)\citenamefont {Wu},
  \citenamefont {Li},\ and\ \citenamefont {Wu}}]{Wu2015}%
  \BibitemOpen
  \bibfield  {author} {\bibinfo {author} {\bibfnamefont {C.-K.}\ \bibnamefont
  {Wu}}, \bibinfo {author} {\bibfnamefont {C.-K.}\ \bibnamefont {Li}}, \ and\
  \bibinfo {author} {\bibfnamefont {Y.-R.}\ \bibnamefont {Wu}},\ }\href
  {\doibase 10.1007/s10825-015-0688-y} {\bibfield  {journal} {\bibinfo
  {journal} {Journal of Computational Electronics}\ }\textbf {\bibinfo {volume}
  {14}},\ \bibinfo {pages} {416} (\bibinfo {year} {2015})}\BibitemShut
  {NoStop}%
\bibitem [{\citenamefont {Filoche}\ and\ \citenamefont
  {Mayboroda}(2012)}]{Filoche2012}%
  \BibitemOpen
  \bibfield  {author} {\bibinfo {author} {\bibfnamefont {M.}~\bibnamefont
  {Filoche}}\ and\ \bibinfo {author} {\bibfnamefont {S.}~\bibnamefont
  {Mayboroda}},\ }\href@noop {} {\bibfield  {journal} {\bibinfo  {journal}
  {Proceedings of the National Academy of Sciences of the USA}\ }\textbf
  {\bibinfo {volume} {109}},\ \bibinfo {pages} {14761} (\bibinfo {year}
  {2012})}\BibitemShut {NoStop}%
\bibitem [{\citenamefont {Arnold}\ \emph {et~al.}(2016)\citenamefont {Arnold},
  \citenamefont {David}, \citenamefont {Jerison}, \citenamefont {Mayboroda},\
  and\ \citenamefont {Filoche}}]{Arnold2016}%
  \BibitemOpen
  \bibfield  {author} {\bibinfo {author} {\bibfnamefont {D.~N.}\ \bibnamefont
  {Arnold}}, \bibinfo {author} {\bibfnamefont {G.}~\bibnamefont {David}},
  \bibinfo {author} {\bibfnamefont {D.}~\bibnamefont {Jerison}}, \bibinfo
  {author} {\bibfnamefont {S.}~\bibnamefont {Mayboroda}}, \ and\ \bibinfo
  {author} {\bibfnamefont {M.}~\bibnamefont {Filoche}},\ }\href@noop {}
  {\bibfield  {journal} {\bibinfo  {journal} {Physical Review Letters}\
  }\textbf {\bibinfo {volume} {116}},\ \bibinfo {pages} {056602} (\bibinfo
  {year} {2016})}\BibitemShut {NoStop}%
\bibitem [{\citenamefont {{De Santi}}\ \emph {et~al.}(2016)\citenamefont {{De
  Santi}}, \citenamefont {Meneghini}, \citenamefont {{La Grassa}},
  \citenamefont {Galler}, \citenamefont {Zeisel}, \citenamefont {Goano},
  \citenamefont {Dominici}, \citenamefont {Mandurrino}, \citenamefont
  {Bertazzi}, \citenamefont {Robidas}, \citenamefont {Meneghesso},\ and\
  \citenamefont {Zanoni}}]{DeSanti2016}%
  \BibitemOpen
  \bibfield  {author} {\bibinfo {author} {\bibfnamefont {C.}~\bibnamefont {{De
  Santi}}}, \bibinfo {author} {\bibfnamefont {M.}~\bibnamefont {Meneghini}},
  \bibinfo {author} {\bibfnamefont {M.}~\bibnamefont {{La Grassa}}}, \bibinfo
  {author} {\bibfnamefont {B.}~\bibnamefont {Galler}}, \bibinfo {author}
  {\bibfnamefont {R.}~\bibnamefont {Zeisel}}, \bibinfo {author} {\bibfnamefont
  {M.}~\bibnamefont {Goano}}, \bibinfo {author} {\bibfnamefont
  {S.}~\bibnamefont {Dominici}}, \bibinfo {author} {\bibfnamefont
  {M.}~\bibnamefont {Mandurrino}}, \bibinfo {author} {\bibfnamefont
  {F.}~\bibnamefont {Bertazzi}}, \bibinfo {author} {\bibfnamefont
  {D.}~\bibnamefont {Robidas}}, \bibinfo {author} {\bibfnamefont
  {G.}~\bibnamefont {Meneghesso}}, \ and\ \bibinfo {author} {\bibfnamefont
  {E.}~\bibnamefont {Zanoni}},\ }\href {\doibase 10.1063/1.4942438} {\bibfield
  {journal} {\bibinfo  {journal} {Journal of Applied Physics}\ }\textbf
  {\bibinfo {volume} {119}},\ \bibinfo {pages} {094501} (\bibinfo {year}
  {2016})},\ \Eprint {http://arxiv.org/abs/http://dx.doi.org/10.1063/1.4942438}
  {http://dx.doi.org/10.1063/1.4942438} \BibitemShut {NoStop}%
\bibitem [{\citenamefont {Zhang}\ \emph {et~al.}(2009)\citenamefont {Zhang},
  \citenamefont {Bhattacharya}, \citenamefont {Singh},\ and\ \citenamefont
  {Hinckley}}]{Zhang2009}%
  \BibitemOpen
  \bibfield  {author} {\bibinfo {author} {\bibfnamefont {M.}~\bibnamefont
  {Zhang}}, \bibinfo {author} {\bibfnamefont {P.}~\bibnamefont {Bhattacharya}},
  \bibinfo {author} {\bibfnamefont {J.}~\bibnamefont {Singh}}, \ and\ \bibinfo
  {author} {\bibfnamefont {J.}~\bibnamefont {Hinckley}},\ }\href {\doibase
  10.1063/1.3266520} {\bibfield  {journal} {\bibinfo  {journal} {Applied
  Physics Letters}\ }\textbf {\bibinfo {volume} {95}},\ \bibinfo {pages}
  {201108} (\bibinfo {year} {2009})}\BibitemShut {NoStop}%
\bibitem [{\citenamefont {David}\ and\ \citenamefont
  {Grundmann}(2010{\natexlab{a}})}]{David2010a}%
  \BibitemOpen
  \bibfield  {author} {\bibinfo {author} {\bibfnamefont {A.}~\bibnamefont
  {David}}\ and\ \bibinfo {author} {\bibfnamefont {M.~J.}\ \bibnamefont
  {Grundmann}},\ }\href {\doibase 10.1063/1.3462916} {\bibfield  {journal}
  {\bibinfo  {journal} {Applied Physics Letters}\ }\textbf {\bibinfo {volume}
  {97}},\ \bibinfo {pages} {033501} (\bibinfo {year}
  {2010}{\natexlab{a}})}\BibitemShut {NoStop}%
\bibitem [{\citenamefont {Kioupakis}\ \emph {et~al.}(2013)\citenamefont
  {Kioupakis}, \citenamefont {Yan}, \citenamefont {Steiauf},\ and\
  \citenamefont {{Van de Walle}}}]{Kioupakis2013}%
  \BibitemOpen
  \bibfield  {author} {\bibinfo {author} {\bibfnamefont {E.}~\bibnamefont
  {Kioupakis}}, \bibinfo {author} {\bibfnamefont {Q.}~\bibnamefont {Yan}},
  \bibinfo {author} {\bibfnamefont {D.}~\bibnamefont {Steiauf}}, \ and\
  \bibinfo {author} {\bibfnamefont {C.~G.}\ \bibnamefont {{Van de Walle}}},\
  }\href@noop {} {\bibfield  {journal} {\bibinfo  {journal} {New Journal of
  Physics}\ }\textbf {\bibinfo {volume} {15}},\ \bibinfo {pages} {125006}
  (\bibinfo {year} {2013})}\BibitemShut {NoStop}%
\bibitem [{\citenamefont {David}\ and\ \citenamefont
  {Grundmann}(2010{\natexlab{b}})}]{David2010c}%
  \BibitemOpen
  \bibfield  {author} {\bibinfo {author} {\bibfnamefont {A.}~\bibnamefont
  {David}}\ and\ \bibinfo {author} {\bibfnamefont {M.~J.}\ \bibnamefont
  {Grundmann}},\ }\href {\doibase 10.1063/1.3330870} {\bibfield  {journal}
  {\bibinfo  {journal} {Applied Physics Letters}\ }\textbf {\bibinfo {volume}
  {96}},\ \bibinfo {pages} {103504} (\bibinfo {year} {2010}{\natexlab{b}})},\
  \Eprint {http://arxiv.org/abs/http://dx.doi.org/10.1063/1.3330870}
  {http://dx.doi.org/10.1063/1.3330870} \BibitemShut {NoStop}%
\bibitem [{\citenamefont {Vaxenburg}\ \emph {et~al.}(2013)\citenamefont
  {Vaxenburg}, \citenamefont {Rodina}, \citenamefont {Lifshitz},\ and\
  \citenamefont {Efros}}]{Vaxenburg2013}%
  \BibitemOpen
  \bibfield  {author} {\bibinfo {author} {\bibfnamefont {R.}~\bibnamefont
  {Vaxenburg}}, \bibinfo {author} {\bibfnamefont {A.}~\bibnamefont {Rodina}},
  \bibinfo {author} {\bibfnamefont {E.}~\bibnamefont {Lifshitz}}, \ and\
  \bibinfo {author} {\bibfnamefont {A.~L.}\ \bibnamefont {Efros}},\ }\href
  {\doibase 10.1063/1.4833915} {\bibfield  {journal} {\bibinfo  {journal}
  {Applied Physics Letters}\ }\textbf {\bibinfo {volume} {103}},\ \bibinfo
  {pages} {221111} (\bibinfo {year} {2013})}\BibitemShut {NoStop}%
\bibitem [{\citenamefont {Lentali}\ \emph {et~al.}(2017)\citenamefont
  {Lentali}, \citenamefont {Piccardo}, \citenamefont {Li}, \citenamefont {Wu},
  \citenamefont {Weisbuch},\ and\ \citenamefont {Filoche}}]{Lentali2017}%
  \BibitemOpen
  \bibfield  {author} {\bibinfo {author} {\bibfnamefont {J.-M.}\ \bibnamefont
  {Lentali}}, \bibinfo {author} {\bibfnamefont {M.}~\bibnamefont {Piccardo}},
  \bibinfo {author} {\bibfnamefont {C.-K.}\ \bibnamefont {Li}}, \bibinfo
  {author} {\bibfnamefont {Y.-R.}\ \bibnamefont {Wu}}, \bibinfo {author}
  {\bibfnamefont {C.}~\bibnamefont {Weisbuch}}, \ and\ \bibinfo {author}
  {\bibfnamefont {M.}~\bibnamefont {Filoche}},\ }\href@noop {} {\bibfield
  {journal} {\bibinfo  {journal} {in preparation}\ } (\bibinfo {year}
  {2017})}\BibitemShut {NoStop}%
\bibitem [{\citenamefont {Anderson}(1958)}]{Anderson1958}%
  \BibitemOpen
  \bibfield  {author} {\bibinfo {author} {\bibfnamefont {P.~W.}\ \bibnamefont
  {Anderson}},\ }\href@noop {} {\bibfield  {journal} {\bibinfo  {journal}
  {Physical Review}\ }\textbf {\bibinfo {volume} {109}},\ \bibinfo {pages}
  {1492} (\bibinfo {year} {1958})}\BibitemShut {NoStop}%
\bibitem [{\citenamefont {Nguyen}\ \emph {et~al.}(2004)\citenamefont {Nguyen},
  \citenamefont {Regnault}, \citenamefont {Ferreira},\ and\ \citenamefont
  {Bastard}}]{Nguyen2004}%
  \BibitemOpen
  \bibfield  {author} {\bibinfo {author} {\bibfnamefont {D.-P.}\ \bibnamefont
  {Nguyen}}, \bibinfo {author} {\bibfnamefont {N.}~\bibnamefont {Regnault}},
  \bibinfo {author} {\bibfnamefont {R.}~\bibnamefont {Ferreira}}, \ and\
  \bibinfo {author} {\bibfnamefont {G.}~\bibnamefont {Bastard}},\ }\href@noop
  {} {\bibfield  {journal} {\bibinfo  {journal} {Solid State Communications}\
  }\textbf {\bibinfo {volume} {130}},\ \bibinfo {pages} {751} (\bibinfo {year}
  {2004})}\BibitemShut {NoStop}%
\bibitem [{\citenamefont {Baranovskii}\ and\ \citenamefont
  {Efros}(1978)}]{Baranovskii1978}%
  \BibitemOpen
  \bibfield  {author} {\bibinfo {author} {\bibfnamefont {S.~D.}\ \bibnamefont
  {Baranovskii}}\ and\ \bibinfo {author} {\bibfnamefont {A.~L.}\ \bibnamefont
  {Efros}},\ }\href@noop {} {\bibfield  {journal} {\bibinfo  {journal} {Soviet
  Physics Semiconductors}\ }\textbf {\bibinfo {volume} {12}},\ \bibinfo {pages}
  {1328} (\bibinfo {year} {1978})}\BibitemShut {NoStop}%
\bibitem [{\citenamefont {Vurgaftman}\ \emph {et~al.}(2001)\citenamefont
  {Vurgaftman}, \citenamefont {Meyer},\ and\ \citenamefont
  {Ram-Mohan}}]{Vurgaftman2001}%
  \BibitemOpen
  \bibfield  {author} {\bibinfo {author} {\bibfnamefont {I.}~\bibnamefont
  {Vurgaftman}}, \bibinfo {author} {\bibfnamefont {J.~R.}\ \bibnamefont
  {Meyer}}, \ and\ \bibinfo {author} {\bibfnamefont {L.~R.}\ \bibnamefont
  {Ram-Mohan}},\ }\href {\doibase 10.1063/1.1368156} {\bibfield  {journal}
  {\bibinfo  {journal} {Journal of Applied Physics}\ }\textbf {\bibinfo
  {volume} {89}},\ \bibinfo {pages} {5815} (\bibinfo {year}
  {2001})}\BibitemShut {NoStop}%
\bibitem [{\citenamefont {Piprek}(2007)}]{Piprek2007}%
  \BibitemOpen
  \bibfield  {author} {\bibinfo {author} {\bibfnamefont {J.}~\bibnamefont
  {Piprek}},\ }\href@noop {} {\emph {\bibinfo {title} {Nitride Semiconductor
  Devices: Principles and Simulation}}}\ (\bibinfo  {publisher} {Wiley},\
  \bibinfo {year} {2007})\BibitemShut {NoStop}%
\bibitem [{\citenamefont {Romanov}\ \emph {et~al.}(2006)\citenamefont
  {Romanov}, \citenamefont {Baker}, \citenamefont {Nakamura},\ and\
  \citenamefont {Speck}}]{Romanov2006}%
  \BibitemOpen
  \bibfield  {author} {\bibinfo {author} {\bibfnamefont {A.~E.}\ \bibnamefont
  {Romanov}}, \bibinfo {author} {\bibfnamefont {T.~J.}\ \bibnamefont {Baker}},
  \bibinfo {author} {\bibfnamefont {S.}~\bibnamefont {Nakamura}}, \ and\
  \bibinfo {author} {\bibfnamefont {J.~S.}\ \bibnamefont {Speck}},\ }\href@noop
  {} {\bibfield  {journal} {\bibinfo  {journal} {Journal of Applied Physics}\
  }\textbf {\bibinfo {volume} {100}} (\bibinfo {year} {2006})}\BibitemShut
  {NoStop}%
\bibitem [{\citenamefont {Fiorentini}\ \emph {et~al.}(2002)\citenamefont
  {Fiorentini}, \citenamefont {Bernardini},\ and\ \citenamefont
  {Ambacher}}]{Fiorentini2002}%
  \BibitemOpen
  \bibfield  {author} {\bibinfo {author} {\bibfnamefont {V.}~\bibnamefont
  {Fiorentini}}, \bibinfo {author} {\bibfnamefont {F.}~\bibnamefont
  {Bernardini}}, \ and\ \bibinfo {author} {\bibfnamefont {O.}~\bibnamefont
  {Ambacher}},\ }\href@noop {} {\bibfield  {journal} {\bibinfo  {journal}
  {Applied Physics Letters}\ }\textbf {\bibinfo {volume} {80}},\ \bibinfo
  {pages} {1204–1206} (\bibinfo {year} {2002})}\BibitemShut {NoStop}%
\bibitem [{\citenamefont {Ambacher}\ \emph {et~al.}(2016)\citenamefont
  {Ambacher}, \citenamefont {Majewski}, \citenamefont {Miskys}, \citenamefont
  {Link}, \citenamefont {Hermann}, \citenamefont {Eickhoff}, \citenamefont
  {Stutzmann}, \citenamefont {Bernardini}, \citenamefont {Fiorentini},
  \citenamefont {Tilak}, \citenamefont {Schaff},\ and\ \citenamefont
  {Eastman}}]{Ambacher2002}%
  \BibitemOpen
  \bibfield  {author} {\bibinfo {author} {\bibfnamefont {O.}~\bibnamefont
  {Ambacher}}, \bibinfo {author} {\bibfnamefont {J.}~\bibnamefont {Majewski}},
  \bibinfo {author} {\bibfnamefont {C.}~\bibnamefont {Miskys}}, \bibinfo
  {author} {\bibfnamefont {A.}~\bibnamefont {Link}}, \bibinfo {author}
  {\bibfnamefont {M.}~\bibnamefont {Hermann}}, \bibinfo {author} {\bibfnamefont
  {M.}~\bibnamefont {Eickhoff}}, \bibinfo {author} {\bibfnamefont
  {M.}~\bibnamefont {Stutzmann}}, \bibinfo {author} {\bibfnamefont
  {F.}~\bibnamefont {Bernardini}}, \bibinfo {author} {\bibfnamefont
  {V.}~\bibnamefont {Fiorentini}}, \bibinfo {author} {\bibfnamefont
  {V.}~\bibnamefont {Tilak}}, \bibinfo {author} {\bibfnamefont
  {B.}~\bibnamefont {Schaff}}, \ and\ \bibinfo {author} {\bibfnamefont {L.~F.}\
  \bibnamefont {Eastman}},\ }\href@noop {} {\bibfield  {journal} {\bibinfo
  {journal} {Journal of Physics: Condensed Matter}\ }\textbf {\bibinfo {volume}
  {14}},\ \bibinfo {pages} {3399} (\bibinfo {year} {2016})}\BibitemShut
  {NoStop}%
\bibitem [{\citenamefont {Hsu}\ \emph {et~al.}(2015)\citenamefont {Hsu},
  \citenamefont {Wu}, \citenamefont {Li}, \citenamefont {Lu},\ and\
  \citenamefont {Wu}}]{Hsu2015}%
  \BibitemOpen
  \bibfield  {author} {\bibinfo {author} {\bibfnamefont {C.~C.}\ \bibnamefont
  {Hsu}}, \bibinfo {author} {\bibfnamefont {C.~K.}\ \bibnamefont {Wu}},
  \bibinfo {author} {\bibfnamefont {C.~K.}\ \bibnamefont {Li}}, \bibinfo
  {author} {\bibfnamefont {T.~C.}\ \bibnamefont {Lu}}, \ and\ \bibinfo {author}
  {\bibfnamefont {Y.~R.}\ \bibnamefont {Wu}},\ }in\ \href {\doibase
  10.1109/NUSOD.2015.7292795} {\emph {\bibinfo {booktitle} {2015 International
  Conference on Numerical Simulation of Optoelectronic Devices (NUSOD)}}}\
  (\bibinfo {year} {2015})\ pp.\ \bibinfo {pages} {7--8}\BibitemShut {NoStop}%
\bibitem [{\citenamefont {Geuzaine}\ and\ \citenamefont
  {Remacle}(2009)}]{Geuzaine2009}%
  \BibitemOpen
  \bibfield  {author} {\bibinfo {author} {\bibfnamefont {C.}~\bibnamefont
  {Geuzaine}}\ and\ \bibinfo {author} {\bibfnamefont {J.~F.}\ \bibnamefont
  {Remacle}},\ }\href@noop {} {\bibfield  {journal} {\bibinfo  {journal}
  {International Journal of Numerical Methods in Engineering}\ }\textbf
  {\bibinfo {volume} {79}},\ \bibinfo {pages} {1309} (\bibinfo {year}
  {2009})}\BibitemShut {NoStop}%
\bibitem [{\citenamefont {Li}\ \emph {et~al.}(2014)\citenamefont {Li},
  \citenamefont {Rosmeulen}, \citenamefont {Simoen},\ and\ \citenamefont
  {Wu}}]{Li2014}%
  \BibitemOpen
  \bibfield  {author} {\bibinfo {author} {\bibfnamefont {C.-K.}\ \bibnamefont
  {Li}}, \bibinfo {author} {\bibfnamefont {M.}~\bibnamefont {Rosmeulen}},
  \bibinfo {author} {\bibfnamefont {E.}~\bibnamefont {Simoen}}, \ and\ \bibinfo
  {author} {\bibfnamefont {Y.-R.}\ \bibnamefont {Wu}},\ }\href {\doibase
  10.1109/TED.2013.2294534} {\bibfield  {journal} {\bibinfo  {journal} {IEEE
  Transactions on Electron Devices}\ }\textbf {\bibinfo {volume} {61}},\
  \bibinfo {pages} {511} (\bibinfo {year} {2014})}\BibitemShut {NoStop}%
\bibitem [{\citenamefont {Kumakura}\ \emph {et~al.}(2005)\citenamefont
  {Kumakura}, \citenamefont {Makimoto}, \citenamefont {Kobayashi},
  \citenamefont {Hashizume}, \citenamefont {Fukui},\ and\ \citenamefont
  {Hasegawa}}]{Kumakura2005}%
  \BibitemOpen
  \bibfield  {author} {\bibinfo {author} {\bibfnamefont {K.}~\bibnamefont
  {Kumakura}}, \bibinfo {author} {\bibfnamefont {T.}~\bibnamefont {Makimoto}},
  \bibinfo {author} {\bibfnamefont {N.}~\bibnamefont {Kobayashi}}, \bibinfo
  {author} {\bibfnamefont {T.}~\bibnamefont {Hashizume}}, \bibinfo {author}
  {\bibfnamefont {T.}~\bibnamefont {Fukui}}, \ and\ \bibinfo {author}
  {\bibfnamefont {H.}~\bibnamefont {Hasegawa}},\ }\href {\doibase
  10.1063/1.1861116} {\bibfield  {journal} {\bibinfo  {journal} {Applied
  Physics Letters}\ }\textbf {\bibinfo {volume} {86}},\ \bibinfo {pages}
  {052105} (\bibinfo {year} {2005})}\BibitemShut {NoStop}%
\bibitem [{\citenamefont {Narukawa}\ \emph {et~al.}(2010)\citenamefont
  {Narukawa}, \citenamefont {Ichikawa}, \citenamefont {Sanga}, \citenamefont
  {Sano},\ and\ \citenamefont {Mukai}}]{Narukawa2010}%
  \BibitemOpen
  \bibfield  {author} {\bibinfo {author} {\bibfnamefont {Y.}~\bibnamefont
  {Narukawa}}, \bibinfo {author} {\bibfnamefont {M.}~\bibnamefont {Ichikawa}},
  \bibinfo {author} {\bibfnamefont {D.}~\bibnamefont {Sanga}}, \bibinfo
  {author} {\bibfnamefont {M.}~\bibnamefont {Sano}}, \ and\ \bibinfo {author}
  {\bibfnamefont {T.}~\bibnamefont {Mukai}},\ }\href {\doibase
  10.1088/0022-3727/43/35/354002} {\bibfield  {journal} {\bibinfo  {journal}
  {Journal of Physics D: Applied Physics}\ }\textbf {\bibinfo {volume} {43}},\
  \bibinfo {pages} {354002} (\bibinfo {year} {2010})}\BibitemShut {NoStop}%
\bibitem [{\citenamefont {Li}\ \emph {et~al.}(2016)\citenamefont {Li},
  \citenamefont {Wu}, \citenamefont {Hsu}, \citenamefont {Lu}, \citenamefont
  {Li}, \citenamefont {Lu},\ and\ \citenamefont {Wu}}]{Li2016}%
  \BibitemOpen
  \bibfield  {author} {\bibinfo {author} {\bibfnamefont {C.-K.}\ \bibnamefont
  {Li}}, \bibinfo {author} {\bibfnamefont {C.-K.}\ \bibnamefont {Wu}}, \bibinfo
  {author} {\bibfnamefont {C.-C.}\ \bibnamefont {Hsu}}, \bibinfo {author}
  {\bibfnamefont {L.-S.}\ \bibnamefont {Lu}}, \bibinfo {author} {\bibfnamefont
  {H.}~\bibnamefont {Li}}, \bibinfo {author} {\bibfnamefont {T.-C.}\
  \bibnamefont {Lu}}, \ and\ \bibinfo {author} {\bibfnamefont {Y.-R.}\
  \bibnamefont {Wu}},\ }\href {\doibase 10.1063/1.4950771} {\bibfield
  {journal} {\bibinfo  {journal} {AIP Advances}\ }\textbf {\bibinfo {volume}
  {6}},\ \bibinfo {pages} {055208} (\bibinfo {year} {2016})}\BibitemShut
  {NoStop}%
\bibitem [{\citenamefont {Fireman}\ \emph {et~al.}(2016)\citenamefont
  {Fireman}, \citenamefont {Browne}, \citenamefont {Mishra},\ and\
  \citenamefont {Speck}}]{Fireman2016}%
  \BibitemOpen
  \bibfield  {author} {\bibinfo {author} {\bibfnamefont {M.~N.}\ \bibnamefont
  {Fireman}}, \bibinfo {author} {\bibfnamefont {D.~A.}\ \bibnamefont {Browne}},
  \bibinfo {author} {\bibfnamefont {U.~K.}\ \bibnamefont {Mishra}}, \ and\
  \bibinfo {author} {\bibfnamefont {J.~S.}\ \bibnamefont {Speck}},\ }\href
  {\doibase 10.1063/1.4941323} {\bibfield  {journal} {\bibinfo  {journal}
  {Journal of Applied Physics}\ }\textbf {\bibinfo {volume} {119}},\ \bibinfo
  {pages} {055709} (\bibinfo {year} {2016})}\BibitemShut {NoStop}%
\bibitem [{\citenamefont {Binder}\ \emph
  {et~al.}(2013{\natexlab{b}})\citenamefont {Binder}, \citenamefont {Galler},
  \citenamefont {Furitsch}, \citenamefont {Off}, \citenamefont {Wagner},
  \citenamefont {Zeisel},\ and\ \citenamefont {Katz}}]{Binder2013b}%
  \BibitemOpen
  \bibfield  {author} {\bibinfo {author} {\bibfnamefont {M.}~\bibnamefont
  {Binder}}, \bibinfo {author} {\bibfnamefont {B.}~\bibnamefont {Galler}},
  \bibinfo {author} {\bibfnamefont {M.}~\bibnamefont {Furitsch}}, \bibinfo
  {author} {\bibfnamefont {J.}~\bibnamefont {Off}}, \bibinfo {author}
  {\bibfnamefont {J.}~\bibnamefont {Wagner}}, \bibinfo {author} {\bibfnamefont
  {R.}~\bibnamefont {Zeisel}}, \ and\ \bibinfo {author} {\bibfnamefont
  {S.}~\bibnamefont {Katz}},\ }\href {\doibase 10.1063/1.4833895} {\bibfield
  {journal} {\bibinfo  {journal} {Applied Physics Letters}\ }\textbf {\bibinfo
  {volume} {103}},\ \bibinfo {pages} {221110} (\bibinfo {year}
  {2013}{\natexlab{b}})},\ \Eprint
  {http://arxiv.org/abs/http://dx.doi.org/10.1063/1.4833895}
  {http://dx.doi.org/10.1063/1.4833895} \BibitemShut {NoStop}%
\bibitem [{\citenamefont {Zhu}\ \emph {et~al.}(2009)\citenamefont {Zhu},
  \citenamefont {Xu}, \citenamefont {Noemaun}, \citenamefont {Kim},
  \citenamefont {Schubert}, \citenamefont {Crawford},\ and\ \citenamefont
  {Koleske}}]{Zhu2009}%
  \BibitemOpen
  \bibfield  {author} {\bibinfo {author} {\bibfnamefont {D.}~\bibnamefont
  {Zhu}}, \bibinfo {author} {\bibfnamefont {J.}~\bibnamefont {Xu}}, \bibinfo
  {author} {\bibfnamefont {A.~N.}\ \bibnamefont {Noemaun}}, \bibinfo {author}
  {\bibfnamefont {J.~K.}\ \bibnamefont {Kim}}, \bibinfo {author} {\bibfnamefont
  {E.~F.}\ \bibnamefont {Schubert}}, \bibinfo {author} {\bibfnamefont {M.~H.}\
  \bibnamefont {Crawford}}, \ and\ \bibinfo {author} {\bibfnamefont {D.~D.}\
  \bibnamefont {Koleske}},\ }\href {\doibase 10.1063/1.3089687} {\bibfield
  {journal} {\bibinfo  {journal} {Applied Physics Letters}\ }\textbf {\bibinfo
  {volume} {94}},\ \bibinfo {pages} {081113} (\bibinfo {year}
  {2009})}\BibitemShut {NoStop}%
\bibitem [{\citenamefont {Sah}\ \emph {et~al.}(1957)\citenamefont {Sah},
  \citenamefont {Noyce},\ and\ \citenamefont {Shockley}}]{Sah1957}%
  \BibitemOpen
  \bibfield  {author} {\bibinfo {author} {\bibfnamefont {C.}~\bibnamefont
  {Sah}}, \bibinfo {author} {\bibfnamefont {R.~N.}\ \bibnamefont {Noyce}}, \
  and\ \bibinfo {author} {\bibfnamefont {W.}~\bibnamefont {Shockley}},\ }\href
  {\doibase 10.1109/JRPROC.1957.278528} {\bibfield  {journal} {\bibinfo
  {journal} {Proceedings of the IRE}\ }\textbf {\bibinfo {volume} {45}},\
  \bibinfo {pages} {1228} (\bibinfo {year} {1957})}\BibitemShut {NoStop}%
\bibitem [{\citenamefont {David}\ \emph {et~al.}(2016)\citenamefont {David},
  \citenamefont {Hurni}, \citenamefont {Young},\ and\ \citenamefont
  {Craven}}]{David2016}%
  \BibitemOpen
  \bibfield  {author} {\bibinfo {author} {\bibfnamefont {A.}~\bibnamefont
  {David}}, \bibinfo {author} {\bibfnamefont {C.~A.}\ \bibnamefont {Hurni}},
  \bibinfo {author} {\bibfnamefont {N.~G.}\ \bibnamefont {Young}}, \ and\
  \bibinfo {author} {\bibfnamefont {M.~D.}\ \bibnamefont {Craven}},\ }\href
  {\doibase 10.1063/1.4961491} {\bibfield  {journal} {\bibinfo  {journal}
  {Applied Physics Letters}\ }\textbf {\bibinfo {volume} {109}},\ \bibinfo
  {pages} {083501} (\bibinfo {year} {2016})}\BibitemShut {NoStop}%
\bibitem [{\citenamefont {David}\ \emph {et~al.}(2008)\citenamefont {David},
  \citenamefont {Grundmann}, \citenamefont {Kaeding}, \citenamefont {Gardner},
  \citenamefont {Mihopoulos},\ and\ \citenamefont {Krames}}]{David2008}%
  \BibitemOpen
  \bibfield  {author} {\bibinfo {author} {\bibfnamefont {A.}~\bibnamefont
  {David}}, \bibinfo {author} {\bibfnamefont {M.~J.}\ \bibnamefont
  {Grundmann}}, \bibinfo {author} {\bibfnamefont {J.~F.}\ \bibnamefont
  {Kaeding}}, \bibinfo {author} {\bibfnamefont {N.~F.}\ \bibnamefont
  {Gardner}}, \bibinfo {author} {\bibfnamefont {T.~G.}\ \bibnamefont
  {Mihopoulos}}, \ and\ \bibinfo {author} {\bibfnamefont {M.~R.}\ \bibnamefont
  {Krames}},\ }\href {\doibase 10.1063/1.2839305} {\bibfield  {journal}
  {\bibinfo  {journal} {Applied Physics Letters}\ }\textbf {\bibinfo {volume}
  {92}},\ \bibinfo {pages} {053502} (\bibinfo {year} {2008})}\BibitemShut
  {NoStop}%
\bibitem [{\citenamefont {Lehoucq}\ \emph {et~al.}(1997)\citenamefont
  {Lehoucq}, \citenamefont {Sorensen},\ and\ \citenamefont
  {Yang}}]{Lehoucq1997}%
  \BibitemOpen
  \bibfield  {author} {\bibinfo {author} {\bibfnamefont {R.~B.}\ \bibnamefont
  {Lehoucq}}, \bibinfo {author} {\bibfnamefont {D.~C.}\ \bibnamefont
  {Sorensen}}, \ and\ \bibinfo {author} {\bibfnamefont {C.}~\bibnamefont
  {Yang}},\ }\href@noop {} {\enquote {\bibinfo {title} {{ARPACK Users Guide:
  Solution of Large Scale Eigenvalue Problems by Implicitly Restarted Arnoldi
  Methods}},}\ } (\bibinfo {year} {1997})\BibitemShut {NoStop}%
\bibitem [{\citenamefont {Schenk}\ \emph {et~al.}(2001)\citenamefont {Schenk},
  \citenamefont {G\"artner}, \citenamefont {Fichtner},\ and\ \citenamefont
  {Stricker}}]{Schenk2001}%
  \BibitemOpen
  \bibfield  {author} {\bibinfo {author} {\bibfnamefont {O.}~\bibnamefont
  {Schenk}}, \bibinfo {author} {\bibfnamefont {K.}~\bibnamefont {G\"artner}},
  \bibinfo {author} {\bibfnamefont {W.}~\bibnamefont {Fichtner}}, \ and\
  \bibinfo {author} {\bibfnamefont {A.}~\bibnamefont {Stricker}},\ }\href@noop
  {} {\bibfield  {journal} {\bibinfo  {journal} {Future Generation Computer
  Systems}\ }\textbf {\bibinfo {volume} {18}},\ \bibinfo {pages} {69} (\bibinfo
  {year} {2001})}\BibitemShut {NoStop}%
\bibitem [{\citenamefont {Watson-Parris}(2011)}]{WatsonParris2011b}%
  \BibitemOpen
  \bibfield  {author} {\bibinfo {author} {\bibfnamefont {D.}~\bibnamefont
  {Watson-Parris}},\ }\emph {\bibinfo {title} {Carrier Localization in
  InGaN/GaN Quantum Wells}},\ \href@noop {} {Ph.D. thesis},\ \bibinfo  {school}
  {University of Manchester} (\bibinfo {year} {2011})\BibitemShut {NoStop}%
\end{thebibliography}%

\end{document}